\documentclass[structabstract]{aa}  

\usepackage{graphicx}
\usepackage{txfonts}
\usepackage{multirow}
\usepackage{aalongtable}

\begin{document}

   \title{GRB 091127: The cooling break race on magnetic fuel}

   \author{R. Filgas\inst{1} \and J. Greiner\inst{1} \and P. Schady\inst{1} 
           \and T. Kr{\"u}hler\inst{1,2,3}
           \and A. C. Updike\inst{4,5,6} \and S. Klose\inst{7} 
           \and M. Nardini\inst{1}\thanks{Present address: Universit\`a degli studi di Milano-Bicocca, Piazza della Scienza 3, 20126, Milano, Italy} 
           \and D. A. Kann\inst{7}
           \and A. Rossi\inst{7} \and V. Sudilovsky\inst{1}
           \and P. M. J. Afonso\inst{1}\thanks{Present address: American River College, Physics \& Astronomy Dpt., 4700 College Oak Drive, Sacramento, CA 95841} 
           \and C. Clemens\inst{1} \and J. Elliott\inst{1} 
           \and A. Nicuesa Guelbenzu\inst{7} \and F. Olivares E.\inst{1} \and A. Rau\inst{1}  
            }

   \institute{Max-Planck-Institut f\"ur extraterrestrische Physik,
     Giessenbachstra\ss{}e 1, 85748 Garching, Germany, \\     
   \email{filgas@mpe.mpg.de}
       \and
     Universe Cluster, Technische Universit\"at M\"unchen, 
     Boltzmannstra\ss{}e 2, 85748 Garching, Germany
        \and
     Dark Cosmology Centre, Niels Bohr Institute, University of Copenhagen, Juliane
     Maries Vej 30, 2100 Copenhagen, Denmark   
        \and
     Department of Physics and Astronomy, Clemson University, Clemson, 
     SC 29634-0978, USA
        \and
     CRESST and the Observational Cosmology Laboratory, NASA/GSFC, Greenbelt, MD 20771, USA
         \and
     Department of Astronomy, University of Maryland, College Park, MD 20742, USA  
         \and
     Th\"uringer Landessternwarte Tautenburg, Sternwarte 5,
     07778 Tautenburg, Germany
        }
      
   \date{Received ? July 2011 / Accepted ? July 2011}

 \abstract
   {}
   {Using high-quality, broad-band afterglow data for GRB 091127, we
    investigate the validity of the synchrotron fireball model for gamma-ray bursts, and
    infer physical parameters of the ultra-relativistic outflow.}
   {We used multi-wavelength (NIR to X-ray) follow-up observations obtained
    with GROND simultaneously in the $g'r'i'z'JH$ filters and the XRT onboard the \emph{Swift}
    satellite in the 0.3 to 10 keV energy range. The resulting afterglow light curve is
    of excellent accuracy with relative photometric errors as low as 1\%, and the
    spectral energy distribution is well-sampled over 5 decades in energy. These data
    present one of the most comprehensive observing campaigns for a single GRB afterglow
    and allow us to test several proposed emission models and outflow characteristics in
    unprecedented detail.}
   {Both the multi-color light curve and the broad-band SED of the afterglow
    of GRB 091127 show evidence of a cooling break moving from high to lower energies. The early 
    light curve is well described by a broken power-law, where
    the initial decay in the  optical/NIR wavelength range is considerably flatter
    than at X-rays. Detailed fitting of the time-resolved SED shows that the break is
    very smooth with a sharpness index of $2.2\pm 0.2$, and evolves towards lower
    frequencies  as a power-law with index $-1.23\pm0.06$. These are the first accurate
    and contemporaneous measurements of both the sharpness of the spectral break and its time evolution.}
   {The measured evolution of the cooling break ($\nu_c \propto t^{\sim
    -1.2}$) is not consistent with the predictions of the standard model,
    wherein $\nu_c \propto t^{\sim -0.5}$ is expected. A possible explanation for the
    observed behavior is a time dependence of the microphysical parameters, in particular the
    fraction of the total energy in the magnetic field $\epsilon_B$. This conclusion provides further evidence
    that the standard fireball model is too simplistic, and time-dependent
    micro-physical parameters may be required to model the growing number of well-sampled
    afterglow light curves.}
   \keywords{gamma rays: bursts - ISM: jets and outflows - X-rays: individuals: GRB 091127}
   \maketitle 
   
\section{Introduction}
   Gamma-ray bursts (GRBs) are among the most energetic explosions in the universe. The leading model for their 
   afterglows is the synchrotron fireball (M\'esz\'aros \& Rees \cite{mezsaros};  Piran \cite{piran};
   M{\'e}sz{\'a}ros \cite{mezsaros2}; Zhang \& M{\'e}sz{\'a}ros \cite{zhang}).
   In this model, the afterglow arises from the synchrotron emission of shock-accelerated electrons in a  
   fireball interacting with the circum-burst medium. The spectral
   energy distribution (SED) of such emission is well modeled by several broken power-laws connected at characteristic 
   break frequencies (Sari et al. \cite{sari}). The model predicts a break in the light curve when the cooling
   frequency ($\nu_c$, the frequency of electrons whose radiative cooling time-scale
   equals the dynamical time of the system) or the characteristic synchrotron frequency ($\nu_m$, 
   peak frequency for the minimal energy of the radiating electrons) passes through the observed bands.
   Such breaks in the light curve have been, however, difficult to identify reliably as the passage of the above frequencies.
   
   With the development of rapid-response telescopes and multi-wavelength instruments, 
   we expected to detect the movement of the break frequencies. However, this movement
   has only possibly been observed directly in the afterglow of GRB~080319B (Racusin et al. \cite{racusin2}).
   Detections of the spectral-break movements in other GRBs were mostly based on the evolution of the 
   GRB afterglow light curves in just one 
   or few filters, where the subtle steepening is visible and is attributed to the passage of the cooling frequency, 
   for example GRB~990510 (Kumar \& Panaitescu \cite{kumar}),
   GRB~030329 (Sato et al. \cite{sato}; Uemura et al. \cite{uemura}), 
   GRB~040924 (Huang et al. \cite{huang}), GRB~041218 (Torii et al. \cite{torii}), GRB~050408 (Kann et al. \cite{kann}), 
   GRB~050502A (Yost et al. \cite{yost}), GRB~060729 (Grupe et al. \cite{grupe}), etc. In some cases, this
   claim is supported by measured spectral evolution. Lipkin et al. (\cite{lipkin}) measured the $B-R$ color
   change in the afterglow of GRB~030329, supporting the theory of the cooling break passage derived
   from the light-curve steepening. Only very few GRBs had coverage in several bands good enough to model 
   the evolution of the afterglow spectrum. In one such rare case, de Ugarte Postigo et al. (\cite{deugarte}) modelled 
   the broad-band SED of the afterglow of GRB~021004 at three distinct epochs, 
   though only the low frequency part of the spectrum shows any evolution. 
   In order to study such spectral evolutions in detail, 
   continuous coverage with high signal-to-noise ratio in several bands simultaneously is required.

   The \emph{Swift} satellite (Gehrels et al. \cite{gehrels}) makes it possible to study the afterglow 
   emission starting at very early times thanks to its rapid slewing capability, a precise localization
   of GRBs with its Burst Alert Telescope (BAT, Barthelmy et al. \cite{barthelmy}),
   and early follow-up with onboard telescopes sensitive at X-ray (XRT, Burrows et al. \cite{burrows})
   and ultraviolet/optical (UVOT, Roming et al. \cite{roming}) wavelengths. 
   Since its launch in 2004, \emph{Swift} has provided many early and well-sampled afterglow light curves
   and X-ray spectra. Blustin et al. (\cite{blustin}) for example fitted broad-band SEDs of 
   the afterglow of GRB~050525A with a cooling break between early optical and X-ray data and 
   with a simple power-law through later epochs, suggesting a spectral evolution. However, such 
   sudden spectral change can sometimes be also attributed to another component with a different electron distribution
   present in the emission at later times (Filgas et al. \cite{filgas}). 

   The most convincing
   measurement of the cooling break movement to this date is the case of the naked-eye burst 
   GRB~080319B (supplementary information in Racusin et al. \cite{racusin2}; Schady et al. in prep.).
   Due to the enormous brightness of this event, these authors were able to fit broad-band SEDs at several
   epochs using \emph{Swift} UVOT and XRT data, as well as a multitude of optical and NIR ground-based data, 
   showing a clear temporal evolution of a break that may be attributed to the cooling break. 
   The previously mentioned examples show that in case of regularly bright GRB afterglows small telescopes cannot 
   provide the accuracy needed for such detailed study. 
   
   The Gamma-Ray burst Optical Near-infrared Detector (GROND, Greiner et al. \cite{greiner1};
   Greiner et al. \cite{greiner2}) at the 2.2~m MPI/ESO telescope at La Silla observatory 
   is capable of providing high-quality, very well-sampled data in seven bands simultaneously
   and therefore opening a new region with respect to data quality and quantity. 
   Such high-precision data allow not only for a detailed study of afterglow 
   light curves (Greiner et al. \cite{greiner080129}, 
   Nardini et al. \cite{nardini}) but also jets of GRBs (Kr\"uhler et al. \cite{kruhler}),
   the dust in their host galaxies (Kr\"uhler et al. \cite{kruhler2}, K\"{u}pc\"{u} Yolda\c{s} et al. \cite{KupcuYoldas},
   Greiner et al. \cite{greiner3}, Kr\"uhler et al. subm.), their 
   redshifts (Greiner et al. \cite{greiner080913}, Kr\"uhler et al. \cite{kruhler3}) and much more.

   Here we provide details of the \emph{Swift} and GROND observations of the 
   afterglow of GRB~091127 and discuss the light curves and SEDs in the context of the fireball model
   thanks to very good energy coverage and sampling of our high-quality data.    
   Throughout the paper, we adopt the convention that the flux density 
   of the GRB afterglow can be described as $F_\nu (t) \propto t^{-\alpha} \nu^{-\beta}$, where $\alpha$ is 
   the temporal and $\beta$ the spectral index. Unless stated otherwise in the text, all reported 
   errors are at 1$\sigma$ confidence level.
   
\section{Observations}
 \subsection{Prompt emission}

   At $T_0 =$ 23:25:45~UT, the \emph{Swift}/BAT was triggered by the long GRB~091127 
   (Troja et al. \cite{troja}). Due to an Earth-limb observing constraint, \emph{Swift}
   could not slew to the target until 53~min after the trigger (Immler \& Troja \cite{immler}).
   The mask-weighted light curve shows three main peaks from $T_0-0.3$ to $T_0+10$~s, peaking at
   $\sim T_0$, $T_0+1.1$~s and at $T_0+7$~s. The measured 
   $T_{90}$ (15-350 keV) is $7.1 \pm 0.2$~s (Stamatikos et al. \cite{stamatikos}).
   The BAT prompt emission spectrum from $T_0-0.4$ to $T_0+7.5$~s is best fitted using a simple power-law model
   with photon index $2.05 \pm 0.07$ and the total fluence in the 15-150 keV energy range is 
   $(9.0 \pm 0.3) \times 10^{-6}$~erg~cm$^{-2}$ (Stamatikos et al. \cite{stamatikos}). We can get a better picture
   of the prompt emission from the instruments with larger energy coverage. Konus-Wind observed the burst
   in the 20 keV - 2 MeV energy range and measured a fluence of $(1.22 \pm 0.06) \times 10^{-5}$~erg~cm$^{-2}$. 
   The time-integrated spectrum of the burst (from $T_0$ to $T_0$+8.4~s) is well fitted by a power-law
   with exponential cutoff model with $\alpha$ = $-1.95\pm0.10$, and $E_{\rm{peak}} = 21.3^{+4}_{-3}$~keV 
   (Golenetskii et al. \cite{golenetskii}). Using a standard concordance cosmology ($H_0 = 71.0$~km~s$^{-1}$~Mpc$^{-1}$, 
   $\Omega_M$ = 0.27, $\Omega_{\Lambda} = 0.73$, Komatsu et al. \cite{komatsu}), and a redshift 
   of $z=0.49$ (Cucchiara et al. \cite{cucchiara}; Th{\"o}ne et al. \cite {thoene}), we calculate the bolometric 
   (1 keV - 10 MeV) energy release of GRB~091127 to be $E_{\rm{iso}} = 1.4 \times 10^{52}$~erg.
   Fermi GBM provides even better energy coverage and the obtained time-averaged spectrum from $T_0+0.002$~s to 
   $T_0+9.984$~s is adequately fit by a Band function (Band et al. \cite{band}) 
   with $E_{\rm{peak}} = 35.5 \pm 1.5$~keV, $\alpha_{\mathrm{prompt}}$ = $-1.26\pm0.07$, 
   and $\beta_{\mathrm{prompt}}$ = $-2.22 \pm 0.02$. The event fluence in the 8 - 1000 keV 
   energy range in this time interval is 
   $(1.92 \pm 0.02) \times 10^{-5}$~erg~cm$^{-2}$ (Goldstein et al., in prep.). This results in the bolometric 
   energy release of $E_{\rm{iso}} = 1.6 \times 10^{52}$~erg, making GRB~091127 consistent within $2\sigma$ with the 
   most updated Amati $E_{\rm{peak}}$ - $E_{\rm{iso}}$ relation (Amati et al. \cite{amati}).

 \subsection{Swift XRT}
   The \emph{Swift}/XRT started observations of the field of GRB~091127 53~min after the trigger 
   (Evans et al. \cite{evans3}). The XRT light 
   curve and spectra were obtained from the XRT repository (Evans et al. \cite{evans}; 
   Evans et al. \cite{evans2}). 
   Spectra were grouped using the grppha task and fitted with the GROND data in XSPEC v12 using 
   $\chi^2$ statistics. The combined optical/X-ray spectral energy distributions were fitted with power-law 
   and broken power-law models and two absorbing columns: one Galactic foreground with a hydrogen 
   column density of $N_H = 2.8\times10^{20}$~cm$^{-2}$ (Kalberla et al. \cite{kalberla}) and another one that is 
   local to the GRB host galaxy at $z=0.49$ (Cucchiara et al. \cite{cucchiara}; Th{\"o}ne et al. \cite {thoene}). 
   Only the latter was allowed to vary in the fits. To investigate the dust 
   reddening in the GRB environment, the zdust model was used, which contains Large and Small Magellanic Clouds 
   (LMC, SMC) and Milky Way (MW) extinction laws from Pei (\cite{pei}).
   The errors of the broad-band SED fits on any single parameter were obtained using the \emph{uncert} command in XSPEC. 
   This calculates the error on the parameter in question while allowing all the other non-frozen parameters in the model to vary.
   
 \begin{figure}[ht]
   \resizebox{\hsize}{!}{\includegraphics{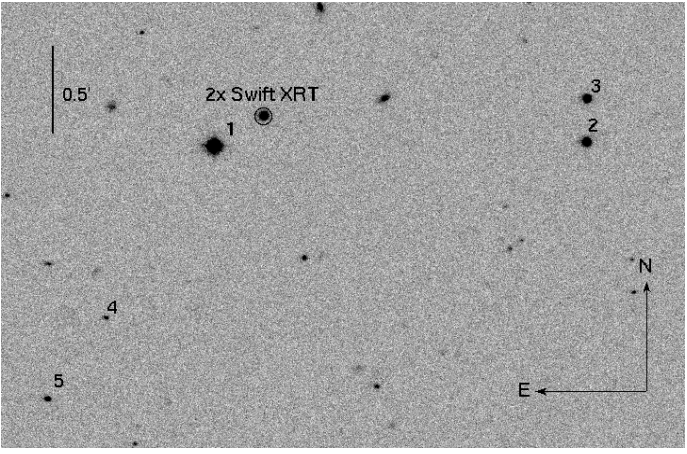}}
   \caption{GROND $r'$ band image of the field of GRB 091127 obtained  
             4.3~ks after T$_0$. The optical afterglow is visible inside the \emph{Swift} XRT
             error circle with double diameter for better clarity. 
             The secondary standard stars are numbered from $1$ to $5$ and
             their magnitudes reported in Table \ref{standards091127}.}
   \label{chart091127}
 \end{figure} 
   
 \subsection{GROND}   
   
   GROND responded to the \emph{Swift} GRB alert and initiated automated
   observations at 00:24~UT, 58~m after the trigger (Updike et al. \cite{updike}).
   GROND imaging of the field of GRB~091127 continued for ten further epochs, the last being acquired
   on October 31st, 2010. Due to the broken chip of the NIR K-band detector, there are no data available
   for this filter. A variable point source was detected in all other bands
   by the automated GROND pipeline (K{\" u}pc{\" u} Yolda{\c s} et al. \cite{yoldas}).
   The position of the transient was calculated to be R.A. (J2000) = 02:26:19.87 and Dec. (J2000) 
   = $-$18:57:08.6 compared to USNO-B reference field stars (Monet et al. 
   \cite{monet}) with an astrometric uncertainty of $0.\!\!^{\prime\prime}3$. 
    
   The optical and NIR image reduction and photometry were performed using standard
   IRAF tasks (Tody \cite{tody}) similar to the procedure described in detail
   in Kr\"uhler et al. (\cite{kruhler2}). A general model for the point-spread function 
   (PSF) of each image was constructed using bright field stars and fitted to the 
   afterglow. In addition, aperture photometry was carried out, and the results
   were consistent with the reported PSF photometry. All data were corrected
   for a Galactic foreground reddening of $E_{\mathrm{B-V}}=0.04$ mag in the direction 
   of the burst (Schlegel et al. \cite{schlegel}), corresponding to an extinction
   of $A_V=0.12$ using $R_V=3.1$, and in the case of $JH$ data, transformed to AB magnitudes.     

   Optical photometric calibration was performed relative to the magnitudes of five secondary 
   standards in the GRB field, shown in Fig. \ref{chart091127} and Table \ref{standards091127}. During 
   photometric conditions, a spectrophotometric standard star SA94-242, a primary
   SDSS standard (Smith et al. \cite{smith}), was observed within a few minutes of observations
   of the GRB field. The obtained zeropoints were corrected
   for atmospheric extinction and used to calibrate stars in the GRB 
   field. The apparent magnitudes 
   of the afterglow were measured with respect to the secondary standards
   reported in Table \ref{standards091127}. The absolute calibration of $JH$ bands
   was obtained with respect to magnitudes of the Two Micron All Sky Survey
   (2MASS) stars within the GRB field obtained from the 2MASS catalog 
   (Skrutskie et al. \cite{skrutskie}). All data are listed
   in Tables \ref{griz} and \ref{JH}.
   
  \begin{figure}[h]
    \resizebox{\hsize}{!}{\includegraphics{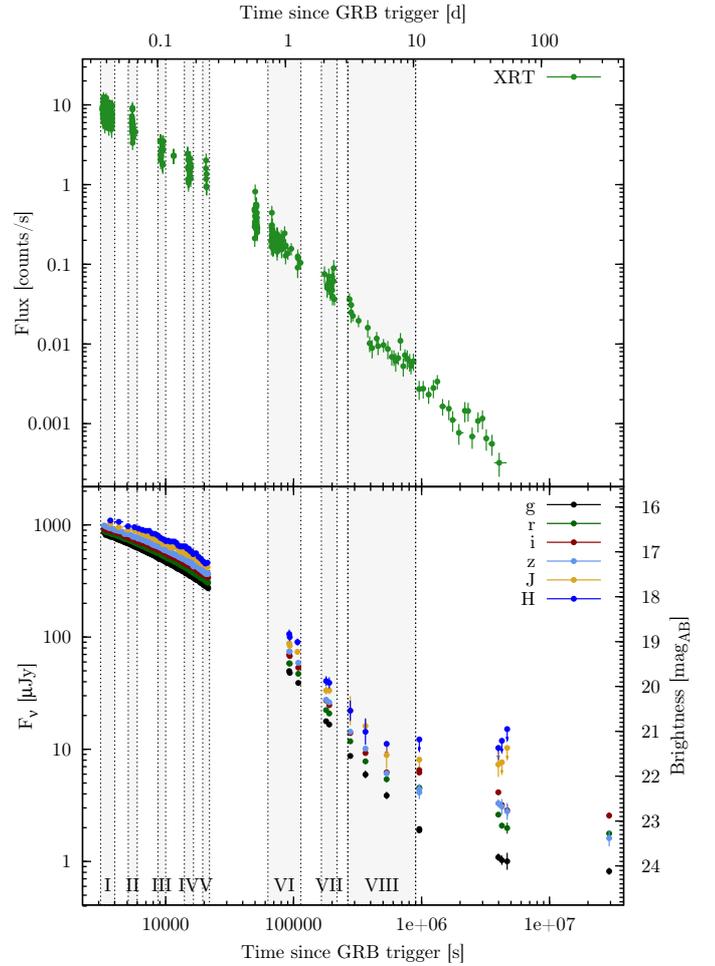}}
    \caption{Light curve of the X-ray (top panel) and GROND optical/NIR (bottom panel)
     afterglow of GRB 091127. Shown data are corrected for Galactic foreground extinction and
       are in AB magnitudes.
       Gray regions show the time intervals where broad-band SEDs were created
      (Fig. \ref{BBfit}).}
    \label{lightcurve}
  \end{figure}    

  \begin{figure}[h]
    \resizebox{\hsize}{!}{\includegraphics[angle=270]{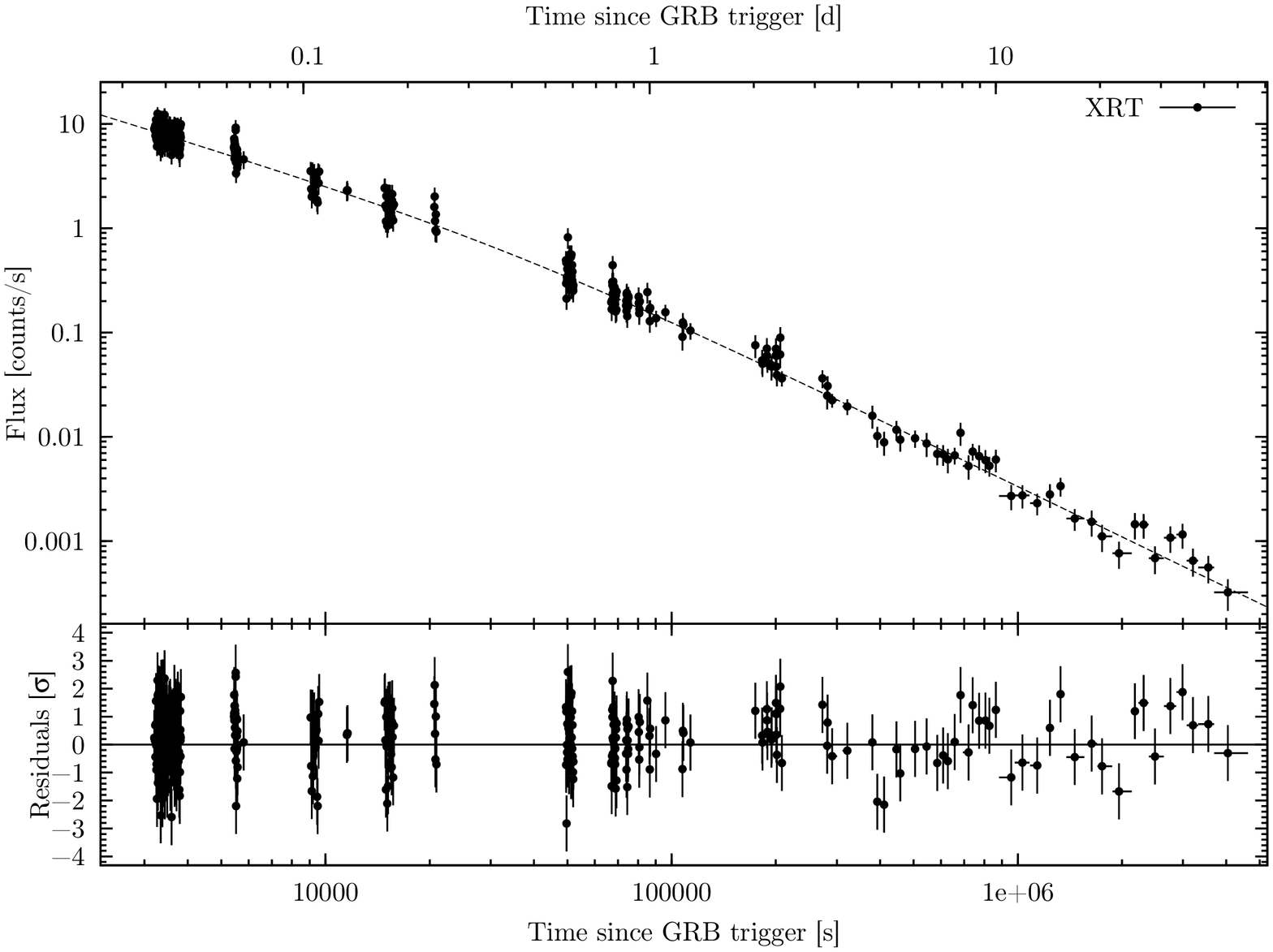}}
    
    \vspace{0.5cm}
    
    \resizebox{\hsize}{!}{\includegraphics[angle=270]{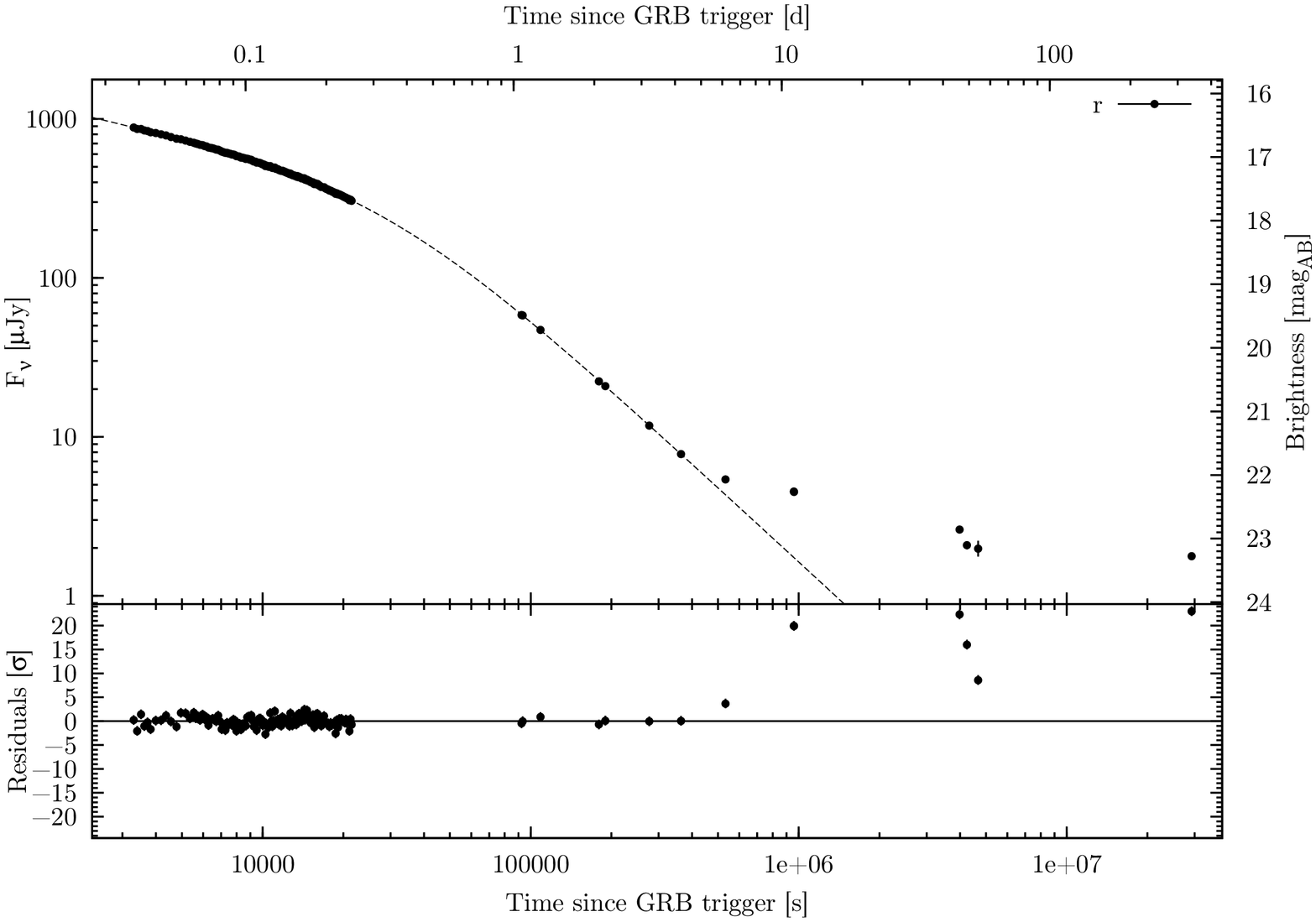}}     
     \caption{The smoothly broken power-law fit to the X-ray light curve (top) and 
      the GROND r' band data (bottom), the parameters of the fit are listed in Table \ref{fittable}.
      Residuals from the best-fit to the $r^\prime$ band data show the SN bump.}
     \label{lightcurvefit}
  \end{figure} 
  
\section{Results}
 \subsection{Afterglow Light Curve}
   
   The X-ray light curve (Fig. \ref{lightcurve}) of the afterglow of GRB~091127 is best fitted 
   with a smoothly broken power-law model 
   (Beuermann et al. \cite{beuermann}) 
   with an initial decay slope $\alpha_X = 1.02\pm0.04$,
   a time of the break at around 33~ks and a post-break temporal slope of $1.61\pm0.04$ 
   (Fig \ref{lightcurvefit}, red. $\chi^2 = 1.03$, straight power-law has red. $\chi^2 = 1.80$, sharply broken
   power-law has red. $\chi^2 = 1.04$).
   The optical/NIR light curve follows the same model but with a much flatter initial temporal slope, which further 
   flattens with increasing wavelength of GROND filters. Table \ref{fittable} shows results of the 
   fitting of a smoothly broken power-law model to each band separately. The sharply broken power-law model
   provides a much worse fit with red. $\chi^2 > 10$ in the optical bands.
   This initial temporal slope is however difficult to measure because the pre-break optical/NIR data show
   a smooth curvature without a straight power-law segment. The reported temporal 
   slope parameters fitted to these data should therefore be considered as estimates of power-law slopes 
   of the earliest optical/NIR data.
 
   The difference in the early decay between X-ray and optical/NIR wavelengths and among optical/NIR bands themselves suggest 
   a strong color evolution, which we discuss in detail in the next section. 
   The time of the X-ray break and the later decay index of the X-ray fit is within 1$\sigma$ errors of
   the fit to the optical bands and within 3$\sigma$ errors of the fit to the NIR bands.
   The optical/NIR data after 500~ks are not fitted as they show contribution from the SN 2009nz bump described 
   by Cobb et al. (\cite{cobb}), Berger et al. (\cite{berger}) and Vergani et al. (\cite{vergani}). 
   We did not subtract the SN magnitudes from 
   the afterglow because this work is based mostly on the early data where the afterglow is dominant.
   Moreover, at even later times, the GROND decay after the break is consistent with the X-ray temporal slope, 
   and the GROND SEDs are well-fitted with a straight power-law. We therefore argue that the influence of 
   the emission not coming from 
   the GRB itself is negligible throughout the time interval used for this study.
    
  \begin{table*}[t]
   \caption{Light curve fit parameters for the afterglow of GRB~091127. The temporal slopes have inaccuracies 
   caused by a very smooth break, which reduces the number of datapoints used in the power-law slopes fitting. 
   The fitting of the NIR bands is affected by the somewhat lower signal-to-noise ratio of the NIR data as compared 
   to the optical bands.}
   \label{fittable}      
   \centering                         
   \begin{tabular}{c c c c c c}        
   \hline\hline    
   \\            
    Band & $\alpha_1$ & $t_{\rm{break}} [s]$ & $s$ & $\alpha_2$ & $\chi^2$/d.o.f.\\    
    \hline 
    XRT & $1.019 \pm 0.039$ & $33472 \pm 3349$ & $2.367 \pm 0.986$ & $1.605 \pm 0.038$ & 373 / 363 \\
    g' & $0.427 \pm 0.011$ & $33917 \pm 2047$ & $1.210 \pm 0.125$ & $1.687 \pm 0.050$ & 125 / 144 \\
    r' & $0.376 \pm 0.009$ & $29287 \pm 1195$ & $1.274 \pm 0.100$ & $1.557 \pm 0.033$ & 143 / 144 \\
    i' & $0.359 \pm 0.014$ & $30288 \pm 1671$ & $1.293 \pm 0.141$ & $1.532 \pm 0.042$ & 133 / 144 \\
    z' & $0.321 \pm 0.016$ & $32368 \pm 2295$ & $1.054 \pm 0.124$ & $1.609 \pm 0.056$ & 131 / 144 \\
    J & $0.300 \pm 0.077$ & $24462 \pm 4453$ & $1.483 \pm 0.728$ & $1.396 \pm 0.147$ & 26 / 37 \\
    H & $0.164 \pm 0.057$ & $21677 \pm 4310$ & $1.005 \pm 0.106$ & $1.417 \pm 0.068$ & 34 / 37 \\
    \hline
   \end{tabular}
   \end{table*}

  \begin{figure}[h]
   \resizebox{\hsize}{!}{\includegraphics{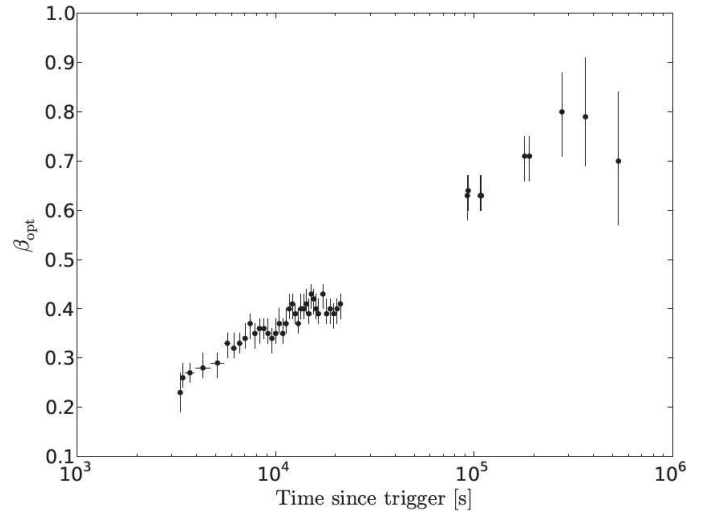}}
   \caption{The optical/NIR spectral slope as a function of time.}
   \label{GRONDsed2}
  \end{figure}
  
 \subsection{Afterglow SEDs}
 
   As already evident from the afterglow light curves, there is a strong spectral evolution in the optical/NIR
   wavelengths before the break.
   Thanks to the simultaneous multi-band observing capabilities of GROND, 
   it is possible to measure the optical/NIR spectral slope as a function of time with high
   accuracy. Fig. \ref{GRONDsed2} shows that the optical/NIR spectral index rises
   from $0.23\pm0.04$ to $0.80\pm0.08$ between 3 and 300~ks. In addition, broad-band optical/NIR to X-ray 
   SEDs were constructed at eight different time intervals within this period, which are indicated in the 
   light curve (Fig. \ref{lightcurve}). Fits of optical/NIR data alone as well as the broad-band fits 
   resulted in a host dust extinction that was consistent with zero, therefore in all the models we 
   assumed no host dust extinction for simplicity.
   
   Fitting the XRT-only spectrum using the full dataset we obtain the host absorbing column density
   $N_H = (1.3 \pm 0.5) \times10^{21}$~cm$^{-2}$.
   Because the broad-band SEDs proved to be inconsistent with a simple power-law model, we used models
   that include a break between the X-ray and optical/NIR data.   
   We initially fitted all eight epochs of broad-band SEDs simultaneously with a sharp broken power-law model,
   where the host-intrinsic absorbing column density and the X-ray spectral index are tied between each SED but left
   free to vary. The low energy spectral indices and energy of the break 
   were left untied between SEDs and free to vary. 
   The best fit (red. $\chi^2 = 1.11$) gives values of the host-equivalent neutral hydrogen 
   density $N_H = (3.2\pm0.6)\times10^{20}$~cm$^{-2}$
   and the high-energy spectral index $\beta_X=0.748\pm0.004$. The value of $N_H$ is smaller than what we get using 
   just the XRT data alone but is consistent within $2\sigma$ with the one resulting from the XRT-only spectral fitting.
   
\begin{figure*}[t]
  \centering
         \includegraphics[height=.95\columnwidth,angle=270]{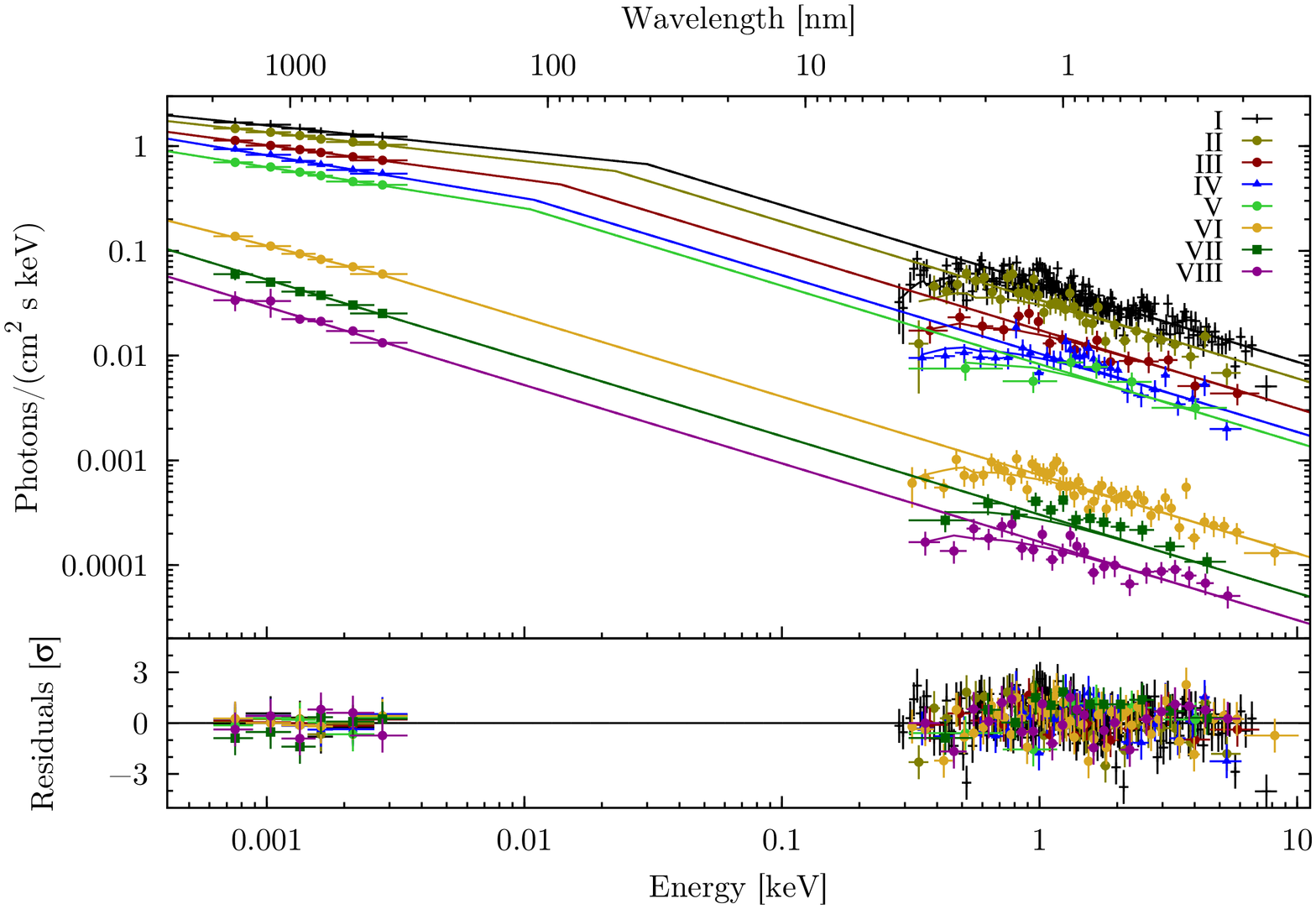}
         \hspace{0.7cm}
         \vspace{0.7cm}
         \includegraphics[height=.9\columnwidth,angle=270]{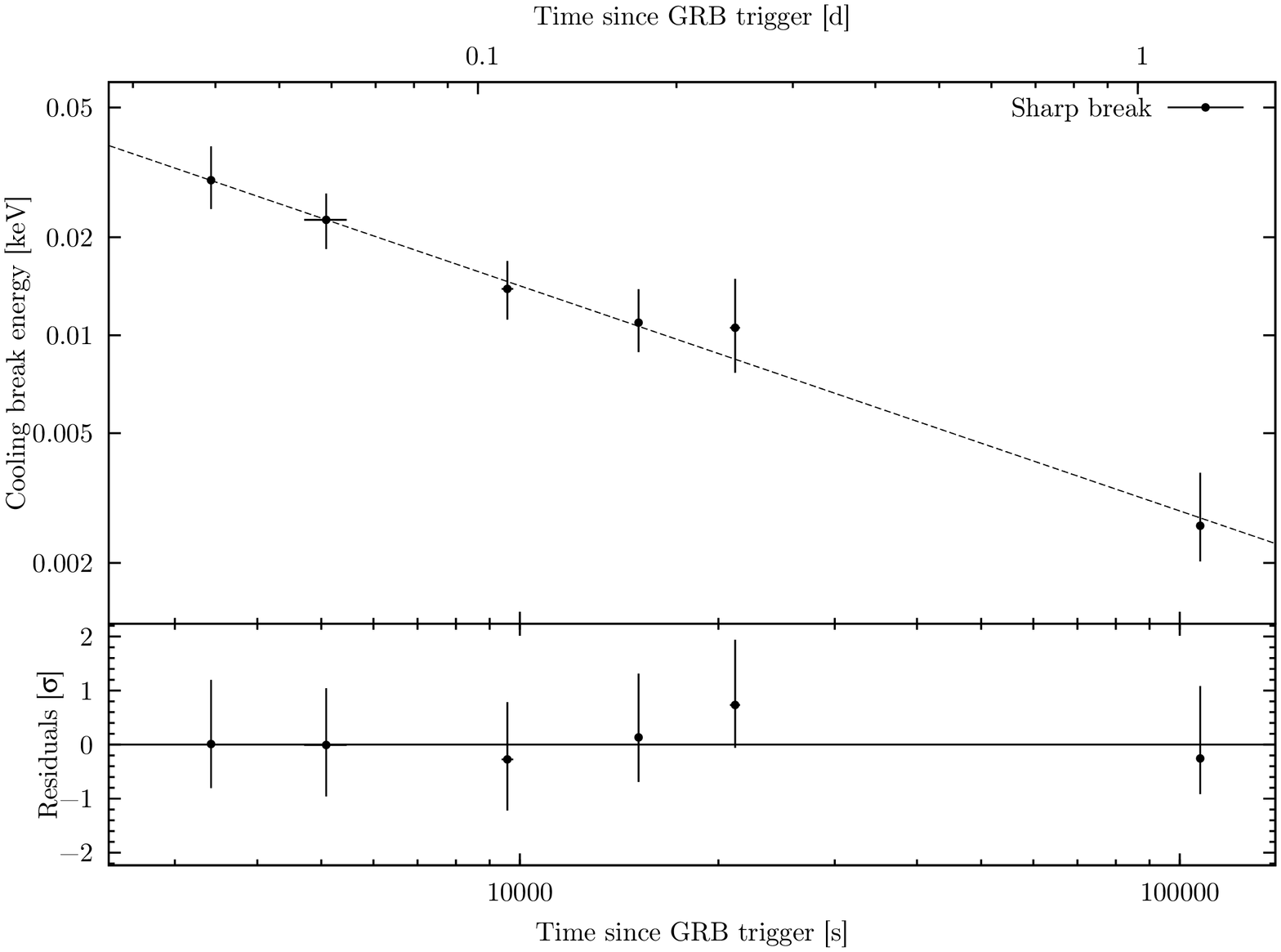} 
         \includegraphics[height=.95\columnwidth,angle=270]{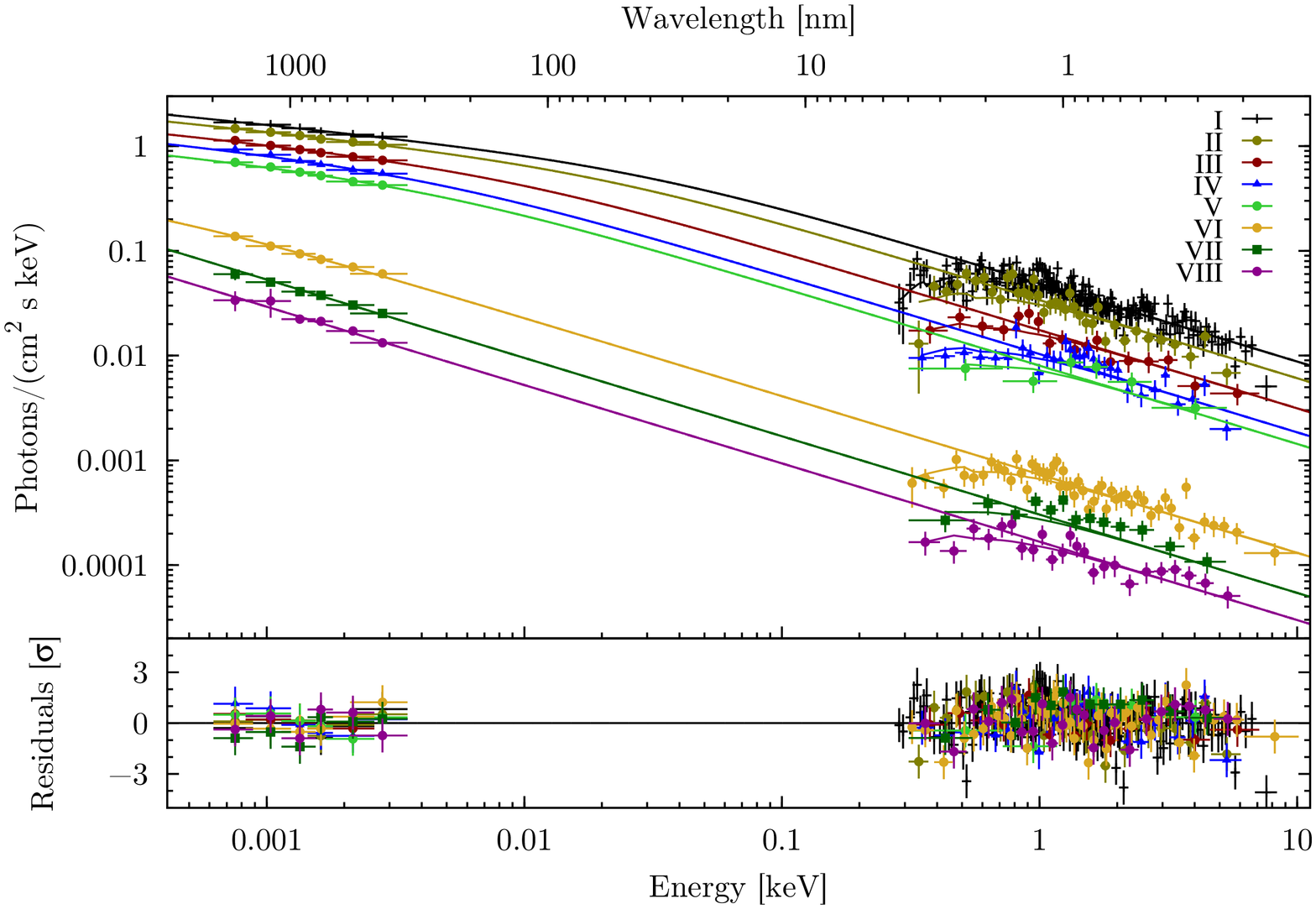}
         \hspace{0.7cm}
         \includegraphics[height=.9\columnwidth,angle=270]{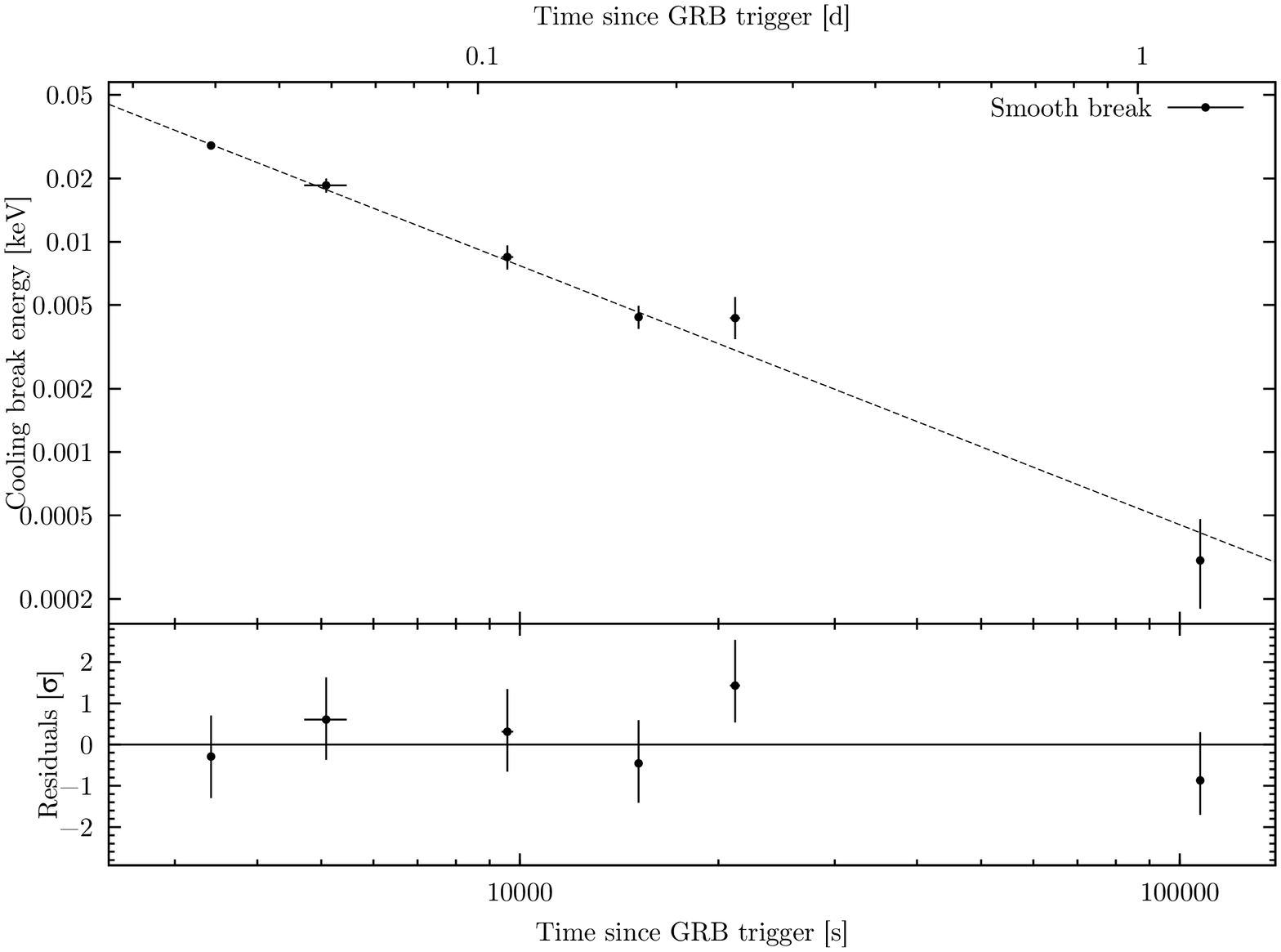}
   \caption{Broad-band optical/NIR to X-ray SEDs fitted with a broken power-law with the sharp break (top left)
        and with a broken power-law with the smooth break (bottom left).
        Best-fit power-law fits to the temporal evolution of the cooling-break energy are shown on the right, 
        resulting from the sharp (top) and the smooth (bottom) broken power-law fits.}
   \label{BBfit} 
 \end{figure*}

   The best-fit optical parameters are listed in Table \ref{BBtable}. 
   This fit shows that the break evolves to larger wavelengths in time, through and
   beyond the optical/NIR bands (top panels of Fig. \ref{BBfit}). 
   The last two SEDs are consistent with a simple power-law continuum without any break. 
   This is in agreement with the X-ray spectral index being within 1$\sigma$ errors consistent with optical/NIR-only
   spectral indices $0.71\pm0.04$ (at time of SED VII) and $0.80\pm0.08$ (at time of SED VIII). The temporal evolution of the 
   break was fitted with a power-law $\nu_c \propto t^x$ and the best-fit index was $x=-0.69 \pm 0.10$ (Fig. \ref{BBfit}).

   Because the fit using the sharp break requires the low-energy spectral index $\beta_{\rm{opt}}$ to be time-dependent,
   we needed a model that would be consistent with constant spectral indices that the theory expects.
   We therefore also fitted all eight broad-band SEDs simultaneously with two power-laws connected by a smooth break
   with flux density following
   \begin{equation}
   F_\nu \propto \left[ \left( {\nu/\nu_{\mathrm{break}}}\right)^{-s\beta_1} + 
   \left( {\nu/\nu_{\mathrm{break}}}\right)^{-s\beta_2}\right] ^{-1/s},
   \label{smootheq}
   \end{equation}
   where $s$ is a parameter that describes the sharpness of the break. 
   Given that the break is far from the X-ray bands, we do not expect the change in the model from a sharp to a smooth 
   break to change the best-fit values of the host absorbing column density $N_H$ nor the high-energy spectral index $\beta_X$. 
   We therefore froze $N_H$ and $\beta_X$ to the best-fit value from  
   the sharp broken power-law fit in order to reduce the number of free parameters in this more complicated model.
   We fixed the difference in values between low and high energy spectral indices to 0.5 (as predicted for the cooling
   break by the standard fireball model; Sari et al. \cite{sari}). The smoothness
   of the break was tied between each SED but left free to vary and the break energy was left free to vary 
   completely. The fit (Fig. \ref{BBfit}, lower panels) again shows the break moving towards the lower energies but in 
   this case the movement is faster than with the sharp break and the fit of the energy over time gives a power-law
   slope of $-1.23 \pm 0.06$.
   
 \subsection{Closure relations} 
   Using the X-ray light-curve fit and the results from the broad-band SEDs, we can test the closure relations
   (Granot \& Sari \cite{granot}; Dai \& Cheng \cite{dai}; Zhang \& M\'esz\'aros \cite{zhang}; 
   Racusin et al. \cite{racusin}) between temporal and spectral indices. The fit-derived X-ray spectral index
   $\beta_X=0.75$ results in a fairly hard power-law index of the electron energy distribution $p = 1.50 \pm 0.01$.
   In the X-rays, the equation (Racusin et al. \cite{racusin}) for $1<p<2$ and a constant decay in the 
   $\nu_X > \nu_\mathrm{c}$ regime, where the jet is interacting with a homogeneous interstellar 
   medium (ISM) and is in the slow cooling phase,
   gives value of $\alpha_X=0.91$ for the spectral index $\beta_X=0.75$ derived from the fits. This value is within 3$\sigma$ 
   of the X-ray light curve pre-break decay slope of $1.02\pm0.04$. However, the fast cooling phase 
   in the $\nu_X > \nu_\mathrm{m}$ regime gives the same value, therefore we cannot distinguish between fast 
   and slow cooling. 
   
   The light curve break at X-rays around 33 ksec must obviously be due to a different phenomenon than the cooling break,
   as the latter started already below the X-ray band at $\sim$3~ks, and then moved to longer wavelengths.
   The post-break evolution of 
   the X-ray light curve is best fitted with the equation describing a non-spreading uniform jet in the ISM, which gives
   $\alpha_X=1.66$, a value consistent within 2$\sigma$ of the fit-derived $1.61 \pm 0.04$. This 
   suggests that, despite the X-ray decay slopes being shallower than the canonical values (Zhang et al. \cite{zhang2};
   Nousek et al. \cite{nousek}), the break in the light curve at around 33~ks represents a jet break 
   (Sari et al. \cite{sari2}). Such shallow ($<2$ with high confidence) post-break decay slopes have been seen in multiple
   well-sampled optical light curves (Zeh et al. \cite{zeh}).
   From the time of the break we can estimate the opening angle of the jet
   to be $\theta \sim 4^\circ$ (Burrows \& Racusin \cite{burrows2}), 
   substituting the measured quantities and normalizing to the typical values $n = 1$~cm$^{-3}$ and $\eta = 0.2$.
   These values lead to the beaming factor and the true gamma-ray energy release (Frail et al. 
   \cite{frail}; Bloom et al. \cite{bloom}) of $f_b = (1-\cos \theta_{jet}) = 2.4 \times 10^{-3}$
   and $E_{\gamma} = 3.9 \times 10^{49}$~erg. For a value of $n=3$~cm$^{-3}$, which is the standard value used
   for the Ghirlanda relation (Ghirlanda et al. \cite{ghirlanda}), we get a jet opening angle 
   $\theta \sim 4.9^\circ$ and $E_{\gamma} = 5.9 \times 10^{49}$~erg. With these values, GRB~091127 lies 
   within the $1\sigma$ scatter of the Ghirlanda relation.
  
  \begin{table*}[t]
   \caption{Best-fit parameters resulting from the sharp and smooth broken power-law fits to the broad-band SEDs. The smoothness
            of the break in the fit using the smooth break between the low- and high-energy spectral index is $2.2 \pm 0.2$.}
   \label{BBtable}      
   \centering                         
   \begin{tabular}{c c c c c}        
   \hline\hline    
   \\            
    SED & Midtime [s] & Low energy spectral & Cooling break [eV] & Cooling break [eV]  \\  
    number & of SED & index using sharp break & using sharp break & using smooth break \\
    \hline \\
    I & 3404 & $0.25^{+0.02}_{-0.04}$ & $29.9 ^{+8.1}_{-5.5}$ & $28.7^{+1.1}_{-1.1}$ \\ [6pt]
    II & 5088 & $0.28 \pm 0.04$ & $22.6^{+4.6}_{-4.2}$ & $18.5^{+1.5}_{-1.4}$\\ [6pt]
    III & 9576 & $0.33^{+0.03}_{-0.04}$ & $13.9^{+3.0}_{-2.7}$  & $8.5^{+1.2}_{-1.1}$ \\ [6pt]
    IV & 15135 & $0.41^{+0.03}_{-0.03}$ & $10.9^{+2.9}_{-2.0}$ & $4.4^{+0.6}_{-0.5}$\\ [6pt]
    V & 21193 & $0.39^{+0.04}_{-0.03}$ & $10.5^{+4.4}_{-2.9}$ & $4.3^{+1.1}_{-0.9}$ \\ [6pt]
    VI & 107401 & $0.62^{+0.04}_{-0.05}$ & $2.6^{+1.2}_{-0.6}$ & $0.3^{+0.2}_{-0.1}$ \\ [6pt]
    VII & 189939 & - & $<0.7$ & $<0.7$ \\ [6pt]
    VIII & 277071 & -  & $<0.7$ &  $<0.7$ \\ 
    \hline
   \end{tabular}
  \end{table*}
  
\section{Discussion}
   The high quality of the data allows us to discuss whether any characteristic synchrotron spectral break could  
   be responsible for the break in the afterglow SED of GRB~091127, and to constrain the sharpness of the break.

 \subsection{Injection break}
   The shape of our broad-band SEDs suggests that the only plausible scenario for the break to be $\nu_m$ is
   the fast-cooling case (Sari et al. \cite{sari}).
   According to the equations in Dai \& Cheng (\cite{dai}), in the case of an ISM medium and for $p=1.5$,
   the characteristic synchrotron frequency $\nu_m$ moves towards lower frequencies as $t^{-2.6}$.
   That is too fast to be consistent with our measurements of the break evolution both for the sharp and the smooth
   break. 
   The predicted light curve slope of $\alpha=0.25$ before the passage
   of the injection break is slightly flatter than our early optical slope. But as previously stated, this
   slope determination is difficult due to the smooth curvature of the early optical/NIR light curve.
   
   However, it is the low-energy spectral slope that is least consistent with the injection break scenario. 
   The SED below $\nu_m$ is expected to be a power-law with index $0.5$, completely independent of 
   the electron energy distribution $p$. This is not consistent with either the sharp break, where the 
   initial slope is a factor 2 flatter and moreover evolving in time, or the smooth break, where the low-energy
   slope is $0.25$ throughout the observation. While this value was fixed in the smooth-break fit, 
   any steeper low-energy slope makes the fit considerably worse and the initial flat optical/NIR only SEDs impossible
   to explain. Therefore we argue that the moving break in the afterglow of GRB~091127 cannot be interpreted
   as the characteristic synchrotron frequency $\nu_m$.

 \subsection{Cooling break}
 \subsubsection{Theoretical expectations}
   According to theory (Sari et al. \cite{sari}; Dai \& Cheng \cite{dai}), 
   in case of an ISM circum-burst environment, the cooling break 
   moves towards lower frequencies with time as a power-law with index $-0.5$. This is within $2\sigma$ 
   of the sharp break fits (Fig. \ref{BBfit}), where the break moves with 
   index $-0.69 \pm 0.10$.
   However, the sharp-break fit requires temporal change of the low-energy spectral index. This is inconsistent
   with the fireball model, where the difference between low- and high-energy spectral indices below and above
   the cooling frequency is constant and $\Delta\beta=0.5$. 
   
   To satisfy the condition of a constant $\Delta\beta$, we fitted the SEDs with a smooth break, 
   that can gradually change the spectral
   index of the data, which occupies a sufficiently narrow portion of the spectra (in this case optical/NIR wavelengths)
   to not show evidence for inherent curvature.
   The smooth-break fit therefore allows both low- and high-energy indices to remain constant, while 
   changing the spectral index fit to GROND data with time, as the break crosses the optical bands (Fig. \ref{GRONDsed2}). 
   Before any further
   discussion, we need to address the question of the physical plausibility of the smooth break.
   
   When we examined the SEDs from studies of large GRB samples (Greiner et al. \cite{greiner3}; 
   Schady et al. \cite{schady1}; Nardini et al. \cite{nardini2}; Schady et al. \cite{schady2}; 
   Starling et al. \cite{starling}), we see
   that they are well fitted with a sharp cooling break (where the break is plausible). 
   This simplistic choice works well for sample studies where it is difficult to distinguish between a sharp 
   and a smooth shape of the break either because the break is far enough from the measured data or 
   because the data lack sufficient
   quality to constrain the smoothness parameter, but can fail in cases like GRB~091127, where extremely large 
   multi-color data sets are available.
   Although previous studies did not require more complex models, Granot \& Sari (\cite{granot})
   calculated that the power-laws in the afterglow spectra are indeed connected by smooth breaks. The theoretical
   smoothness of the cooling break is $1.15-0.06p = 1.06$ for $p=2 \times \beta_X=1.5$. 
   This is roughly a factor of 2 less (i.e., smoother) than our fit-derived smoothness of $2.2 \pm 0.2$. 
   
   The significant inconsistency, however, is related to the speed of the cooling break, which in the smooth fit
   moves with an index $-1.23 \pm 0.06$, a value much higher than the expected $-0.5$. 
   Similar to the value of $-1.00 \pm 0.14$ 
   derived for the cooling break movement reported by Racusin et al. (\cite{racusin}), it would require that we abandon some  
   simplifications often assumed in the simplest formulations of the fireball model. 
   The flux evolution for adiabatic slow cooling in this synchrotron emission theory 
   is described by Eq. (8) in Sari et al. (\cite{sari}) and for convenience we report it here as
   \begin{equation}
   F_\nu = \left\{ 
    \begin{array}{ll}
    {\left(\nu/\nu_m \right)}^{-(p-1)/2} F_{\nu,\mathrm{max}}, & \nu_c > \nu > \nu_m, \\ [8pt]
    {(\nu_c/\nu_m)}^{-(p-1)/2} {(\nu/\nu_c)}^{-p/2} F_{\nu,\mathrm{max}}, & \nu > \nu_c,
       \end{array} \right.
   \label{sarieq}
   \end{equation}
   where the break frequencies for the case of $p<2$ can be calculated from Dai \& Cheng (\cite{dai}) and
   Chevalier \& Li (\cite{chevalier}) to be
   \begin{equation}
   \begin{array}{l}
    \nu_c \propto \epsilon_B^{-3/2} E_{\rm{iso}}^{-1/2} t^{-1/2} , \\ [8pt]
    \nu_m \propto \epsilon_B^{1/{2(p-1)}} \epsilon_e^{2/(p-1)} E_{\rm{iso}}^{p+2/{8(p-1)}} t^{-3(p+2)/{8(p-1)}} , \\ [8pt]
    F_{\nu,\mathrm{max}} \propto \epsilon_B^{1/2} E_{\rm{iso}} ,
       \end{array} 
   \label{sarieq2}
   \end{equation}
   where $t$ is the time since the GRB trigger, $E_{\rm{iso}}$ is the isotropic energy of the GRB, $\epsilon_B$ 
   is the fraction of the energy carried by the magnetic field and $\epsilon_e$ the fraction of the energy in 
   electrons. 
   In the standard fireball model, all parameters are constant in time and the 
   density in the ISM is homogeneous. For the cooling break speed to be 
   consistent with our measurements, one of the parameters $\epsilon_B$ and $E_{\rm{iso}}$ (or a combination
   of them) must evolve with time. Using Eq. \ref{sarieq} and \ref{sarieq2}, we can easily examine 
   cases where each of these parameters evolves separately and model the impact of such an evolution on the 
   resulting afterglow flux.
  
\subsubsection{Theoretical implications} 
   To obtain the measured cooling break speed of $t^{-1.23\pm0.06}$ we need one of the parameters (we treat them 
   separately for simplicity) to add $t^{-0.73\pm0.06}$ to the theoretical speed of $t^{-0.5}$. As we can
   see from the Eq. \ref{sarieq}, the change of the flux evolution before and after the cooling break passage
   is proportional to the cooling break frequency evolution as $\nu_c^{0.5}$. This means that the cooling
   break that is faster by a factor of $t^{-0.73\pm0.06}$ would add $\Delta\alpha=0.37\pm0.03$ to the standard
   change of the temporal index of $\Delta\alpha=0.25$ (Sari et al. \cite{sari}) caused by the cooling 
   brake passage. 

   As we already stated, 
   the early optical/NIR slope is difficult to obtain. However, we can estimate it by calculating the weighted mean 
   of the values of the optical/NIR parameter $\alpha_1$ in Table \ref{fittable}. This results in a decay index of
   $\alpha=0.38$ before the jet break 
   at around 33~ks. If we assume this to be the decay index before the cooling break passage, and we take the X-ray 
   pre-jet-break temporal slope of $\alpha=1.02\pm0.04$ to be the one after the cooling break passage, we get 
   a very good (within $1\sigma$) consistency with our calculated $\Delta\alpha=0.62\pm0.03$. While the 
   amount by which the light-curve steepens is only dependent on the speed of the cooling break and not
   on which parameter causes it, the flux evolution and therefore the decay index itself before and
   after the cooling break passage depends strongly on which parameter we let evolve in time. 
   Using Eq. \ref{sarieq} and \ref{sarieq2}, we can calculate how the time evolution of the flux density 
   depends on these parameters for $p<2$ (for $p>2$ see Eq. B7 and B8 in Panaitescu \& Kumar \cite{panaitescu2}).
   We calculate
   
   \begin{equation}
   F_\nu \propto \left\{ 
    \begin{array}{ll}
    E^{(p+18)/16} \epsilon_B^{3/4} t^{-3(p+2)/16}, & \nu_c > \nu > \nu_m, \\ [8pt]
    E^{(p+14)/16} t^{-(3p+10)/16}, & \nu > \nu_c.
     \end{array} \right.
   \label{myeq}
   \end{equation}   
      
   Letting the isotropic energy vary in time results in $F_\nu \propto E^{(p+18)/16}$ for $\nu<\nu_c$
   and $F_\nu \propto E^{(p+14)/16}$ for $\nu>\nu_c$. In this case the increased speed of the cooling break
   is the result of the isotropic energy which increases in time as $t^{1.46}$. This dependence using
   the fit-derived $p=1.5$ decreases the temporal index before and after the cooling break passage by 
   $1.78$ and $1.41$ respectively. Such extreme flattening of the 
   light curve would mean that without the energy injection the decay slope before the jet break 
   would be $\alpha_1=1.02+1.41 =2.4$ and the late temporal slope after the jet break $\alpha_2=1.61+1.41=3.0$,
   values which are unusually steep for a GRB afterglow (Racusin et al. \cite{racusin}). 
   The energy $E_{\rm{iso}}$ is directly dependent on the energy injection and indirectly on the density
   profile around the burst and we can examine the influence of the time evolution of these parameters on 
   the energy using equations from Sari \& M{\'e}sz{\'a}ros (\cite{sari3}).
   
   The density profile of the medium can be calculated from the cooling-break 
   temporal exponent using equations in Table 1 of Sari \& M{\'e}sz{\'a}ros (\cite{sari3}).
   There $\nu_c \propto t^{(3g-4)/2(4-g)}$, where g is the power-law index of the external density 
   profile $n \propto r^{-g}$. The same approach was used by Racusin et al. (\cite{racusin}) for GRB~080319B
   where the cooling break speed of $t^{-1}$ results in the steep density profile $n \propto r^4$, which 
   requires the existence of a complex medium with a density enhancement. However, our cooling break 
   speed of $t^{-1.23}$ implies an implausibly steep density profile of $n \propto r^{11}$, which would be
   very difficult to defend physically and support observationally.

   Using Eq. 11 in Sari et al. (\cite{sari})
   for the cooling break frequency and assuming typical values of $n_1=1$ and $\epsilon_B=0.01$, we can 
   calculate the isotropic energy of the burst at times corresponding to the first (SED I) and the last (SED VI)
   point where we measure the position of the cooling break using the smooth break fit. The best-fit
   parameters in Table \ref{BBtable}
   give $E_{52}\sim 3.8$ at $t=3.4$~ks and $E_{52}\sim1080$ at $t=107.4$~ks. The increasing energy of
   GRBs can possibly be explained by refreshed shocks, where the central engine ejects shells with a range 
   of Lorentz factors. 
   When the slower material catches up with the decelerating ejecta, it re-energizes it 
   (Sari \& M{\'e}sz{\'a}ros \cite{sari3}). However, assuming a constant density profile, 
   this scenario requires extreme energy 
   injection, leading to an injection parameter $s=8.6$ (see Table 1 in Sari \& M{\'e}sz{\'a}ros \cite{sari3}).
   Such a scenario is very unlikely, as it would require the initial low-energy ejecta to be re-energized by a 
   very large amount of energy stored in slowly moving material. 
   It would also require a gradual and continuous energy injection over the time of our light curve
   coverage, i.e. $\sim$10$^6$ sec, a scenario which so far has never been advocated. We therefore
   also consider a change of energy input an unlikely explanation for the temporal behavior of GRB~091127.  
   
   The last option is to let the microphysical parameter $\epsilon_B$ vary in time. To be consistent with our
   measurement of the cooling break speed, the fraction of energy in the magnetic field 
   would have to rise in time as $\epsilon_B \propto t^{0.49}$. Such an evolution would influence the flux
   as $F_\nu \propto \epsilon_B^{3/4}$ for $\nu<\nu_c$ while the flux density is independent of $\epsilon_B$ for $\nu>\nu_c$.
   Therefore the temporal index before the cooling break passage would decrease by $0.37$ on top of the theoretical 
   flux density evolution. 
   This flattening of the temporal index in the $\nu<\nu_c$
   regime would explain the early shallow optical/NIR decay, while the late data after the jet break would not be 
   influenced by an evolving $\epsilon_B$. We can again use Eq. 11 in Sari et al. (\cite{sari}) to estimate 
   the value of $\epsilon_B$, assuming $E_{52}=1.6$ and $n_1=1$. The calculation results in $\epsilon_B=0.013$
   at $t=3.4$~ks, a value consistent with standard models, and $\epsilon_B=0.088$ at $t=107.4$~ks. 
   
   There is a growing number of studies which have modelled broad-band GRB light curves, and these have yielded
   results for $\epsilon_B$ which span several orders of magnitude between different GRBs, with values 
   from $\sim 10^{-5}$ to $\sim 10^{-1}$ 
   (Panaitescu \& Kumar \cite{panaitescu3}; Panaitescu \& Kumar \cite{panaitescu4}; Yost et al. \cite{yost2}), 
   raising questions whether the assumption of $\epsilon_B$ being constant in the simplest fireball model is consistent
   with the observations. Lately, the idea of $\epsilon_B$ increasing in time as a power-law has been discussed and is 
   receiving increasing support from observational data (e.g., Panaitescu et al. \cite{panaitescu}; Kong et al. \cite{kong}).
   There is also the possibility that all the parameters that influence the cooling frequency 
   vary in time simultaneously. However, it would require more sophisticated theoretical work 
   to derive some estimates or constraints on the ratios between them; our data cannot provide such constraints.
   
   The discussion so far was based on the assumption that the environment around the burst is the undisturbed ISM,
   i.e. the radial density profile is constant.
   While this assumption is supported by the closure relations and the direction of the spectral 
   break, we must consider also the possibility that the circum-burst density has a wind profile.
   In that case we would expect from the theory the cooling break to move towards shorter wavelengths 
   as $\nu_c \propto t^{0.5}$. To be consistent with our measurement of $t^{-1.23}$, the parameters
   in Eq. \ref{sarieq2} would have to increase in time so rapidly, that they would effectively reverse 
   the direction of the cooling break movement. Given that we concluded that the time evolution of parameter $E$ 
   is too dramatic in the ISM scenario, the even more rapid increase required here is more unlikely. 
   To reverse the cooling break movement, $\epsilon_B$ would have to increase its time evolution 
   to $t^{1.15}$. While we cannot completely rule out this option due to the inability to compute the exact
   values of $\epsilon_B$ in evolving density, we believe that such rapid time evolution 
   would be difficult to defend against the ISM scenario.
       
\section{Conclusions}
   Since the launch of the \emph{Swift} satellite, there is growing evidence that
   the radiative mechanism responsible for the optical to X-ray GRB emission is not as simple and well  
   understood as previously believed. 
   The growing number of well-sampled data sets (Covino et al. \cite{covino}; 	
   Guidorzi et al. \cite{guidorzi}; Th{\"o}ne et al. \cite{thoene2}; Filgas et al. \cite{filgas}) 
   is beginning to place strong constraints on the fireball model and possible alternatives (e.g., Dar \& De Rujula \cite{dar};
   Dado et al. \cite{dado}). 
   Most GRBs have complex light curves, for which the optical and X-ray 
   emission are seemingly decoupled, thus providing an indication that they are produced by different mechanisms.
   The afterglow of GRB~091127 is one of the few examples in which the light-curve evolution in the
   optical/NIR and X-ray wavelengths is well represented by a broken power-law and, in addition, 
   both light curves show a break at roughly the same time and similar decay slopes after that break. 
   This observational evidence, together with the fact that the optical/NIR to X-ray SED at late times is well
   represented by a single component, leads us to an assumption that the emission in both energy bands has been
   produced by the same radiative mechanism and that this mechanism could be the standard external shock 
   synchrotron radiation.
   
   We observe a clear break in the light curve at around 33~ks, which we interpret as a jet break, based 
   on the fact that it is achromatic and the post-break evolution of all bands is similar. 
   The GROND SEDs show a strong color evolution with the optical/NIR spectral index rising from roughly 0.25
   to 0.75, while the X-ray spectral slope stays constant. The broad-band NIR to X-ray SEDs were
   fitted with a broken power-law with the break moving in time towards larger wavelengths.
   Because the difference between the low- and high-energy spectral index reaches $0.5$ asymptotically, 
   we interpret the spectral break as the cooling break, decreasing in energy with time, as the forward
   shock moves into an ISM-like circumburst medium.
   Since it takes almost all the follow-up time for the optical/NIR spectral slope to gradually steepen from 
   the initial value to the value consistent with the X-ray spectral index, we conclude that the cooling
   break is very smooth in frequency space. 

   The measured cooling break speed of $\nu_c \propto t^{-1.23\pm0.06}$ is faster than expected for a shock
   evolving in a constant density medium and requires
   that one of the parameters that influence the afterglow flux density evolves with time.
   We conclude that the required changes in the energy release $E_{\rm{iso}}$ alone would be too dramatic to be 
   physically plausible and that the most feasible explanation 
   is the evolution of microphysical parameters. Assuming $\epsilon_B$ (the fraction of the energy carried by the 
   magnetic field) to be the only varying parameter, then during the time interval that we measure the position 
   of the cooling break, between 3 and 107~ks, it would rise in time as $\epsilon_B \propto t^{0.49}$, 
   and would reach values of 0.01 and 0.09 at those times, respectively. 
    
   Currently, a complete understanding of the microphysical processes is 
   still lacking. Nonetheless, data from instruments like \emph{Swift} and GROND can
   shed some light on the shock physics. A larger study of the observational data of bursts similar to GRB~091127 
   is necessary to investigate how commonly such changes in $\epsilon_B$ occur in GRB afterglows. Theoretical 
   studies would be warranted to investigate effects which would change $\epsilon_B$ as the fireball expands 
   into its surrounding environment.
   
\begin{acknowledgements}
We thank the anonymous referee for constructive comments that helped to improve
the paper.\\
Part of the funding for GROND (both hardware as well as 
personnel) was generously granted from the Leibniz-Prize to Prof. G. 
Hasinger (DFG grant HA 1850/28-1).\\
This work made use of data supplied by the UK Swift Science 
Data Centre at the University of Leicester.\\
TK acknowledges support by the DFG cluster of excellence Origin and 
Structure of the Universe.\\
TK acknowledges support by the European Commission under the Marie Curie
Intra-European Fellowship Programme.\\
The Dark Cosmology Centre is funded by the Danish National Research Foundation.\\
FOE acknowledges funding of his Ph.D. through the \emph{Deutscher Akademischer Austausch-Dienst} (DAAD)\\
SK, DAK and ANG acknowledge support by DFG grant Kl 766/16-1.\\
AR acknowledges support from the BLANCEFLOR Boncompagni-Ludovisi, n\'ee Bildt foundation.\\
MN acknowledges support by DFG grant SA 2001/2-1.\\
PS acknowledges support by DFG grant SA 2001/1-1.\\
ACU, ANG, DAK and AR are grateful for travel funding support through MPE.
\end{acknowledgements}

\newpage
\begin{longtable}{r c c c c c}
\caption{\label{griz} $g'r'i'z'$ photometric data} \\            
\hline\hline   
$T_\mathrm{mid} - T_0$ [ks] & Exposure [s] & \multicolumn{4}{c}{
Brightness$^{\mathrm{(a)}}$ mag$_\mathrm{AB}$} \\
 & & $g'$ & $r'$ & $i'$ & $z'$ \\
\hline 
\endfirsthead
\caption{continued.}\\
\hline\hline
$T_\mathrm{mid} - T_0$ [ks] & Exposure [s] & \multicolumn{4}{c}{
Brightness$^{\mathrm{(a)}}$ mag$_\mathrm{AB}$} \\
 & & $g'$ & $r'$ & $i'$ & $z'$ \\
\hline
\endhead
\hline
3.3031	&	35	&$	16.57	\pm	0.01	$&$	16.53	\pm	0.01	$&$	16.48	\pm	0.01	$&$	16.41	\pm	0.01	$\\
3.4039	&	35	&$	16.62	\pm	0.01	$&$	16.56	\pm	0.01	$&$	16.51	\pm	0.01	$&$	16.44	\pm	0.01	$\\
3.5195	&	35	&$	16.63	\pm	0.01	$&$	16.56	\pm	0.01	$&$	16.51	\pm	0.01	$&$	16.45	\pm	0.01	$\\
3.6192	&	35	&$	16.65	\pm	0.01	$&$	16.58	\pm	0.01	$&$	16.52	\pm	0.01	$&$	16.47	\pm	0.01	$\\
3.7202	&	35	&$	16.67	\pm	0.01	$&$	16.59	\pm	0.01	$&$	16.53	\pm	0.01	$&$	16.47	\pm	0.01	$\\
3.8222	&	35	&$	16.67	\pm	0.01	$&$	16.61	\pm	0.01	$&$	16.55	\pm	0.01	$&$	16.49	\pm	0.01	$\\
3.9920	&	115	&$	16.69	\pm	0.01	$&$	16.62	\pm	0.01	$&$	16.55	\pm	0.01	$&$	16.51	\pm	0.01	$\\
4.1762	&	115	&$	16.72	\pm	0.01	$&$	16.64	\pm	0.01	$&$	16.58	\pm	0.01	$&$	16.53	\pm	0.01	$\\
4.3638	&	115	&$	16.74	\pm	0.01	$&$	16.66	\pm	0.01	$&$	16.59	\pm	0.01	$&$	16.54	\pm	0.01	$\\
4.5513	&	115	&$	16.76	\pm	0.01	$&$	16.69	\pm	0.01	$&$	16.61	\pm	0.01	$&$	16.55	\pm	0.01	$\\
4.7693	&	115	&$	16.79	\pm	0.01	$&$	16.71	\pm	0.01	$&$	16.65	\pm	0.01	$&$	16.60	\pm	0.01	$\\
4.9561	&	115	&$	16.81	\pm	0.01	$&$	16.72	\pm	0.01	$&$	16.66	\pm	0.01	$&$	16.60	\pm	0.01	$\\
5.1544	&	115	&$	16.82	\pm	0.01	$&$	16.74	\pm	0.01	$&$	16.69	\pm	0.01	$&$	16.61	\pm	0.01	$\\
5.3507	&	115	&$	16.84	\pm	0.01	$&$	16.76	\pm	0.01	$&$	16.69	\pm	0.01	$&$	16.62	\pm	0.01	$\\
5.5330	&	35	&$	16.87	\pm	0.01	$&$	16.77	\pm	0.01	$&$	16.72	\pm	0.01	$&$	16.66	\pm	0.01	$\\
5.6328	&	35	&$	16.86	\pm	0.01	$&$	16.79	\pm	0.01	$&$	16.71	\pm	0.01	$&$	16.66	\pm	0.01	$\\
5.7340	&	35	&$	16.89	\pm	0.01	$&$	16.79	\pm	0.01	$&$	16.73	\pm	0.01	$&$	16.66	\pm	0.01	$\\
5.8360	&	35	&$	16.89	\pm	0.01	$&$	16.81	\pm	0.01	$&$	16.76	\pm	0.01	$&$	16.67	\pm	0.01	$\\
5.9714	&	35	&$	16.91	\pm	0.01	$&$	16.81	\pm	0.01	$&$	16.74	\pm	0.01	$&$	16.67	\pm	0.01	$\\
6.0722	&	35	&$	16.91	\pm	0.01	$&$	16.83	\pm	0.01	$&$	16.75	\pm	0.01	$&$	16.71	\pm	0.01	$\\
6.1740	&	35	&$	16.91	\pm	0.01	$&$	16.83	\pm	0.01	$&$	16.75	\pm	0.01	$&$	16.70	\pm	0.01	$\\
6.2755	&	35	&$	16.93	\pm	0.01	$&$	16.85	\pm	0.01	$&$	16.77	\pm	0.01	$&$	16.70	\pm	0.01	$\\
6.4035	&	35	&$	16.94	\pm	0.01	$&$	16.86	\pm	0.01	$&$	16.80	\pm	0.01	$&$	16.72	\pm	0.01	$\\
6.5029	&	35	&$	16.95	\pm	0.01	$&$	16.86	\pm	0.01	$&$	16.79	\pm	0.01	$&$	16.72	\pm	0.01	$\\
6.6042	&	35	&$	16.95	\pm	0.01	$&$	16.87	\pm	0.01	$&$	16.79	\pm	0.01	$&$	16.74	\pm	0.01	$\\
6.7059	&	35	&$	16.98	\pm	0.01	$&$	16.88	\pm	0.01	$&$	16.81	\pm	0.01	$&$	16.74	\pm	0.01	$\\
6.8294	&	35	&$	16.97	\pm	0.01	$&$	16.89	\pm	0.01	$&$	16.82	\pm	0.01	$&$	16.74	\pm	0.01	$\\
6.9322	&	35	&$	16.99	\pm	0.01	$&$	16.90	\pm	0.01	$&$	16.82	\pm	0.01	$&$	16.75	\pm	0.01	$\\
7.0341	&	35	&$	17.00	\pm	0.01	$&$	16.92	\pm	0.01	$&$	16.83	\pm	0.01	$&$	16.77	\pm	0.01	$\\
7.1359	&	35	&$	17.01	\pm	0.01	$&$	16.92	\pm	0.01	$&$	16.84	\pm	0.01	$&$	16.77	\pm	0.01	$\\
7.2592	&	35	&$	17.02	\pm	0.01	$&$	16.93	\pm	0.01	$&$	16.84	\pm	0.01	$&$	16.78	\pm	0.01	$\\
7.3600	&	35	&$	17.02	\pm	0.01	$&$	16.93	\pm	0.01	$&$	16.85	\pm	0.01	$&$	16.78	\pm	0.01	$\\
7.4623	&	35	&$	17.03	\pm	0.01	$&$	16.94	\pm	0.01	$&$	16.87	\pm	0.01	$&$	16.79	\pm	0.01	$\\
7.5638	&	35	&$	17.04	\pm	0.01	$&$	16.95	\pm	0.01	$&$	16.86	\pm	0.01	$&$	16.80	\pm	0.01	$\\
7.6804	&	35	&$	17.05	\pm	0.01	$&$	16.96	\pm	0.01	$&$	16.88	\pm	0.01	$&$	16.79	\pm	0.01	$\\
7.7819	&	35	&$	17.06	\pm	0.01	$&$	16.96	\pm	0.01	$&$	16.88	\pm	0.01	$&$	16.82	\pm	0.01	$\\
7.8837	&	35	&$	17.07	\pm	0.01	$&$	16.97	\pm	0.01	$&$	16.91	\pm	0.01	$&$	16.84	\pm	0.01	$\\
7.9858	&	35	&$	17.07	\pm	0.01	$&$	16.99	\pm	0.01	$&$	16.89	\pm	0.01	$&$	16.83	\pm	0.01	$\\
8.1035	&	35	&$	17.07	\pm	0.01	$&$	16.99	\pm	0.01	$&$	16.92	\pm	0.01	$&$	16.85	\pm	0.01	$\\
8.2033	&	35	&$	17.09	\pm	0.01	$&$	17.00	\pm	0.01	$&$	16.91	\pm	0.01	$&$	16.85	\pm	0.01	$\\
8.3060	&	35	&$	17.10	\pm	0.01	$&$	17.01	\pm	0.01	$&$	16.92	\pm	0.01	$&$	16.85	\pm	0.01	$\\
8.4083	&	35	&$	17.10	\pm	0.01	$&$	17.01	\pm	0.01	$&$	16.93	\pm	0.01	$&$	16.87	\pm	0.01	$\\
8.5548	&	35	&$	17.13	\pm	0.01	$&$	17.02	\pm	0.01	$&$	16.93	\pm	0.01	$&$	16.87	\pm	0.01	$\\
8.6551	&	35	&$	17.12	\pm	0.01	$&$	17.03	\pm	0.01	$&$	16.93	\pm	0.01	$&$	16.87	\pm	0.01	$\\
8.7563	&	35	&$	17.14	\pm	0.01	$&$	17.03	\pm	0.01	$&$	16.96	\pm	0.01	$&$	16.87	\pm	0.01	$\\
8.8578	&	35	&$	17.14	\pm	0.01	$&$	17.03	\pm	0.01	$&$	16.96	\pm	0.01	$&$	16.89	\pm	0.01	$\\
8.9765	&	35	&$	17.15	\pm	0.01	$&$	17.04	\pm	0.01	$&$	16.95	\pm	0.01	$&$	16.89	\pm	0.01	$\\
9.0752	&	35	&$	17.16	\pm	0.01	$&$	17.04	\pm	0.01	$&$	16.96	\pm	0.01	$&$	16.91	\pm	0.01	$\\
9.1764	&	35	&$	17.17	\pm	0.01	$&$	17.06	\pm	0.01	$&$	16.97	\pm	0.01	$&$	16.91	\pm	0.01	$\\
9.2780	&	35	&$	17.17	\pm	0.01	$&$	17.07	\pm	0.01	$&$	16.97	\pm	0.01	$&$	16.91	\pm	0.01	$\\
9.3991	&	35	&$	17.17	\pm	0.01	$&$	17.07	\pm	0.01	$&$	16.99	\pm	0.01	$&$	16.92	\pm	0.01	$\\
9.4984	&	35	&$	17.18	\pm	0.01	$&$	17.09	\pm	0.01	$&$	17.00	\pm	0.01	$&$	16.92	\pm	0.01	$\\
9.6003	&	35	&$	17.19	\pm	0.01	$&$	17.09	\pm	0.01	$&$	17.01	\pm	0.01	$&$	16.95	\pm	0.01	$\\
9.7024	&	35	&$	17.19	\pm	0.01	$&$	17.09	\pm	0.01	$&$	17.01	\pm	0.01	$&$	16.95	\pm	0.01	$\\
9.8230	&	35	&$	17.21	\pm	0.01	$&$	17.10	\pm	0.01	$&$	17.02	\pm	0.01	$&$	16.96	\pm	0.01	$\\
9.9224	&	35	&$	17.21	\pm	0.01	$&$	17.11	\pm	0.01	$&$	17.02	\pm	0.01	$&$	16.96	\pm	0.01	$\\
10.0238	&	35	&$	17.22	\pm	0.01	$&$	17.12	\pm	0.01	$&$	17.03	\pm	0.01	$&$	16.96	\pm	0.01	$\\
10.1261	&	35	&$	17.22	\pm	0.01	$&$	17.12	\pm	0.01	$&$	17.03	\pm	0.01	$&$	16.97	\pm	0.01	$\\
10.2507	&	35	&$	17.24	\pm	0.01	$&$	17.14	\pm	0.01	$&$	17.03	\pm	0.01	$&$	16.97	\pm	0.01	$\\
10.3520	&	35	&$	17.24	\pm	0.01	$&$	17.14	\pm	0.01	$&$	17.05	\pm	0.01	$&$	16.97	\pm	0.01	$\\
10.4540	&	35	&$	17.25	\pm	0.01	$&$	17.15	\pm	0.01	$&$	17.06	\pm	0.01	$&$	16.99	\pm	0.01	$\\
10.5581	&	35	&$	17.25	\pm	0.01	$&$	17.15	\pm	0.01	$&$	17.08	\pm	0.01	$&$	16.99	\pm	0.01	$\\
10.6782	&	35	&$	17.25	\pm	0.01	$&$	17.14	\pm	0.01	$&$	17.06	\pm	0.01	$&$	17.01	\pm	0.01	$\\
10.7768	&	35	&$	17.26	\pm	0.01	$&$	17.16	\pm	0.01	$&$	17.07	\pm	0.01	$&$	17.00	\pm	0.01	$\\
10.8783	&	35	&$	17.26	\pm	0.01	$&$	17.17	\pm	0.01	$&$	17.06	\pm	0.01	$&$	17.02	\pm	0.01	$\\
10.9796	&	35	&$	17.29	\pm	0.01	$&$	17.17	\pm	0.01	$&$	17.10	\pm	0.01	$&$	17.00	\pm	0.01	$\\
11.1013	&	35	&$	17.28	\pm	0.01	$&$	17.16	\pm	0.01	$&$	17.08	\pm	0.01	$&$	17.03	\pm	0.01	$\\
11.2013	&	35	&$	17.29	\pm	0.01	$&$	17.18	\pm	0.01	$&$	17.10	\pm	0.01	$&$	17.02	\pm	0.01	$\\
11.3023	&	35	&$	17.30	\pm	0.01	$&$	17.19	\pm	0.01	$&$	17.11	\pm	0.01	$&$	17.04	\pm	0.01	$\\
11.4035	&	35	&$	17.29	\pm	0.01	$&$	17.19	\pm	0.01	$&$	17.12	\pm	0.01	$&$	17.03	\pm	0.01	$\\
11.5335	&	35	&$	17.31	\pm	0.01	$&$	17.21	\pm	0.01	$&$	17.11	\pm	0.01	$&$	17.04	\pm	0.01	$\\
11.6330	&	35	&$	17.32	\pm	0.01	$&$	17.21	\pm	0.01	$&$	17.11	\pm	0.01	$&$	17.06	\pm	0.01	$\\
11.7365	&	35	&$	17.33	\pm	0.01	$&$	17.22	\pm	0.01	$&$	17.13	\pm	0.01	$&$	17.06	\pm	0.01	$\\
11.8405	&	35	&$	17.32	\pm	0.01	$&$	17.21	\pm	0.01	$&$	17.14	\pm	0.01	$&$	17.06	\pm	0.01	$\\
11.9610	&	35	&$	17.33	\pm	0.01	$&$	17.22	\pm	0.01	$&$	17.13	\pm	0.01	$&$	17.07	\pm	0.01	$\\
12.0624	&	35	&$	17.34	\pm	0.01	$&$	17.23	\pm	0.01	$&$	17.13	\pm	0.01	$&$	17.07	\pm	0.01	$\\
12.1642	&	35	&$	17.35	\pm	0.01	$&$	17.24	\pm	0.01	$&$	17.14	\pm	0.01	$&$	17.08	\pm	0.01	$\\
12.2655	&	35	&$	17.37	\pm	0.01	$&$	17.24	\pm	0.01	$&$	17.15	\pm	0.01	$&$	17.08	\pm	0.01	$\\
12.3911	&	35	&$	17.35	\pm	0.01	$&$	17.25	\pm	0.01	$&$	17.15	\pm	0.01	$&$	17.08	\pm	0.01	$\\
12.4924	&	35	&$	17.37	\pm	0.01	$&$	17.25	\pm	0.01	$&$	17.16	\pm	0.01	$&$	17.09	\pm	0.01	$\\
12.5943	&	35	&$	17.38	\pm	0.01	$&$	17.27	\pm	0.01	$&$	17.16	\pm	0.01	$&$	17.11	\pm	0.01	$\\
12.6953	&	35	&$	17.37	\pm	0.01	$&$	17.26	\pm	0.01	$&$	17.16	\pm	0.01	$&$	17.09	\pm	0.01	$\\
12.8208	&	35	&$	17.38	\pm	0.01	$&$	17.28	\pm	0.01	$&$	17.17	\pm	0.01	$&$	17.11	\pm	0.01	$\\
12.9194	&	35	&$	17.39	\pm	0.01	$&$	17.28	\pm	0.01	$&$	17.17	\pm	0.01	$&$	17.13	\pm	0.01	$\\
13.0215	&	35	&$	17.39	\pm	0.01	$&$	17.29	\pm	0.01	$&$	17.20	\pm	0.01	$&$	17.11	\pm	0.01	$\\
13.1243	&	35	&$	17.40	\pm	0.01	$&$	17.29	\pm	0.01	$&$	17.19	\pm	0.01	$&$	17.12	\pm	0.01	$\\
13.2482	&	35	&$	17.39	\pm	0.01	$&$	17.29	\pm	0.01	$&$	17.19	\pm	0.01	$&$	17.12	\pm	0.01	$\\
13.3501	&	35	&$	17.42	\pm	0.01	$&$	17.31	\pm	0.01	$&$	17.21	\pm	0.01	$&$	17.14	\pm	0.01	$\\
13.4522	&	35	&$	17.42	\pm	0.01	$&$	17.30	\pm	0.01	$&$	17.21	\pm	0.01	$&$	17.15	\pm	0.01	$\\
13.5537	&	35	&$	17.41	\pm	0.01	$&$	17.31	\pm	0.01	$&$	17.22	\pm	0.01	$&$	17.15	\pm	0.01	$\\
13.6766	&	35	&$	17.43	\pm	0.01	$&$	17.31	\pm	0.01	$&$	17.22	\pm	0.01	$&$	17.14	\pm	0.01	$\\
13.7783	&	35	&$	17.43	\pm	0.01	$&$	17.32	\pm	0.01	$&$	17.22	\pm	0.01	$&$	17.16	\pm	0.01	$\\
13.8797	&	35	&$	17.44	\pm	0.01	$&$	17.32	\pm	0.01	$&$	17.24	\pm	0.01	$&$	17.16	\pm	0.01	$\\
13.9818	&	35	&$	17.44	\pm	0.01	$&$	17.34	\pm	0.01	$&$	17.26	\pm	0.01	$&$	17.15	\pm	0.01	$\\
14.1025	&	35	&$	17.46	\pm	0.01	$&$	17.34	\pm	0.01	$&$	17.25	\pm	0.01	$&$	17.17	\pm	0.01	$\\
14.2035	&	35	&$	17.46	\pm	0.01	$&$	17.34	\pm	0.01	$&$	17.26	\pm	0.01	$&$	17.18	\pm	0.01	$\\
14.3064	&	35	&$	17.47	\pm	0.01	$&$	17.34	\pm	0.01	$&$	17.25	\pm	0.01	$&$	17.16	\pm	0.01	$\\
14.4076	&	35	&$	17.47	\pm	0.01	$&$	17.36	\pm	0.01	$&$	17.25	\pm	0.01	$&$	17.19	\pm	0.01	$\\
14.5296	&	35	&$	17.48	\pm	0.01	$&$	17.36	\pm	0.01	$&$	17.26	\pm	0.01	$&$	17.19	\pm	0.01	$\\
14.6315	&	35	&$	17.47	\pm	0.01	$&$	17.36	\pm	0.01	$&$	17.27	\pm	0.01	$&$	17.19	\pm	0.01	$\\
14.7306	&	35	&$	17.48	\pm	0.01	$&$	17.37	\pm	0.01	$&$	17.27	\pm	0.01	$&$	17.21	\pm	0.01	$\\
14.8318	&	35	&$	17.48	\pm	0.01	$&$	17.37	\pm	0.01	$&$	17.28	\pm	0.01	$&$	17.22	\pm	0.01	$\\
14.9555	&	35	&$	17.49	\pm	0.01	$&$	17.38	\pm	0.01	$&$	17.28	\pm	0.01	$&$	17.21	\pm	0.01	$\\
15.0579	&	35	&$	17.50	\pm	0.01	$&$	17.39	\pm	0.01	$&$	17.31	\pm	0.01	$&$	17.22	\pm	0.01	$\\
15.1596	&	35	&$	17.51	\pm	0.01	$&$	17.40	\pm	0.01	$&$	17.29	\pm	0.01	$&$	17.22	\pm	0.01	$\\
15.2643	&	35	&$	17.50	\pm	0.01	$&$	17.40	\pm	0.01	$&$	17.30	\pm	0.01	$&$	17.23	\pm	0.01	$\\
15.3892	&	35	&$	17.52	\pm	0.01	$&$	17.40	\pm	0.01	$&$	17.30	\pm	0.01	$&$	17.23	\pm	0.01	$\\
15.4886	&	35	&$	17.54	\pm	0.01	$&$	17.42	\pm	0.01	$&$	17.32	\pm	0.01	$&$	17.24	\pm	0.01	$\\
15.5903	&	35	&$	17.53	\pm	0.01	$&$	17.43	\pm	0.01	$&$	17.30	\pm	0.01	$&$	17.23	\pm	0.01	$\\
15.6915	&	35	&$	17.52	\pm	0.01	$&$	17.41	\pm	0.01	$&$	17.30	\pm	0.01	$&$	17.24	\pm	0.01	$\\
15.8137	&	35	&$	17.54	\pm	0.01	$&$	17.42	\pm	0.01	$&$	17.32	\pm	0.01	$&$	17.25	\pm	0.01	$\\
15.9162	&	35	&$	17.54	\pm	0.01	$&$	17.43	\pm	0.01	$&$	17.34	\pm	0.01	$&$	17.26	\pm	0.01	$\\
16.0176	&	35	&$	17.54	\pm	0.01	$&$	17.42	\pm	0.01	$&$	17.32	\pm	0.01	$&$	17.24	\pm	0.01	$\\
16.1197	&	35	&$	17.54	\pm	0.01	$&$	17.44	\pm	0.01	$&$	17.33	\pm	0.01	$&$	17.26	\pm	0.01	$\\
16.2395	&	35	&$	17.57	\pm	0.01	$&$	17.44	\pm	0.01	$&$	17.34	\pm	0.01	$&$	17.27	\pm	0.01	$\\
16.3388	&	35	&$	17.57	\pm	0.01	$&$	17.45	\pm	0.01	$&$	17.32	\pm	0.01	$&$	17.29	\pm	0.01	$\\
16.4424	&	35	&$	17.57	\pm	0.01	$&$	17.46	\pm	0.01	$&$	17.33	\pm	0.01	$&$	17.31	\pm	0.01	$\\
16.5444	&	35	&$	17.58	\pm	0.01	$&$	17.47	\pm	0.01	$&$	17.35	\pm	0.01	$&$	17.33	\pm	0.01	$\\
16.9464	&	115	&$	17.60	\pm	0.01	$&$	17.48	\pm	0.01	$&$	17.37	\pm	0.01	$&$	17.31	\pm	0.01	$\\
17.1273	&	115	&$	17.62	\pm	0.01	$&$	17.49	\pm	0.01	$&$	17.40	\pm	0.01	$&$	17.31	\pm	0.01	$\\
17.3279	&	115	&$	17.62	\pm	0.01	$&$	17.50	\pm	0.01	$&$	17.40	\pm	0.01	$&$	17.32	\pm	0.01	$\\
17.5383	&	115	&$	17.62	\pm	0.01	$&$	17.51	\pm	0.01	$&$	17.41	\pm	0.01	$&$	17.32	\pm	0.01	$\\
17.7360	&	115	&$	17.62	\pm	0.01	$&$	17.53	\pm	0.01	$&$	17.42	\pm	0.01	$&$	17.34	\pm	0.01	$\\
17.9339	&	115	&$	17.63	\pm	0.01	$&$	17.53	\pm	0.01	$&$	17.42	\pm	0.01	$&$	17.34	\pm	0.01	$\\
18.1312	&	115	&$	17.65	\pm	0.01	$&$	17.54	\pm	0.01	$&$	17.45	\pm	0.01	$&$	17.36	\pm	0.01	$\\
18.3158	&	115	&$	17.65	\pm	0.01	$&$	17.55	\pm	0.01	$&$	17.43	\pm	0.01	$&$	17.38	\pm	0.01	$\\
18.5209	&	115	&$	17.67	\pm	0.01	$&$	17.56	\pm	0.01	$&$	17.47	\pm	0.01	$&$	17.38	\pm	0.01	$\\
18.7128	&	115	&$	17.69	\pm	0.01	$&$	17.58	\pm	0.01	$&$	17.44	\pm	0.01	$&$	17.39	\pm	0.01	$\\
18.9061	&	115	&$	17.68	\pm	0.01	$&$	17.57	\pm	0.01	$&$	17.46	\pm	0.01	$&$	17.39	\pm	0.01	$\\
19.1020	&	115	&$	17.69	\pm	0.01	$&$	17.59	\pm	0.01	$&$	17.47	\pm	0.01	$&$	17.42	\pm	0.01	$\\
19.3182	&	115	&$	17.71	\pm	0.01	$&$	17.59	\pm	0.01	$&$	17.50	\pm	0.01	$&$	17.40	\pm	0.01	$\\
19.5012	&	115	&$	17.70	\pm	0.01	$&$	17.60	\pm	0.01	$&$	17.51	\pm	0.01	$&$	17.43	\pm	0.01	$\\
19.6840	&	115	&$	17.73	\pm	0.01	$&$	17.60	\pm	0.01	$&$	17.50	\pm	0.01	$&$	17.42	\pm	0.01	$\\
19.8676	&	115	&$	17.74	\pm	0.01	$&$	17.61	\pm	0.01	$&$	17.51	\pm	0.01	$&$	17.42	\pm	0.01	$\\
20.0781	&	115	&$	17.73	\pm	0.01	$&$	17.62	\pm	0.01	$&$	17.52	\pm	0.01	$&$	17.47	\pm	0.01	$\\
20.2726	&	115	&$	17.75	\pm	0.01	$&$	17.63	\pm	0.01	$&$	17.51	\pm	0.01	$&$	17.46	\pm	0.01	$\\
20.4570	&	115	&$	17.76	\pm	0.01	$&$	17.64	\pm	0.01	$&$	17.52	\pm	0.01	$&$	17.46	\pm	0.01	$\\
20.6577	&	115	&$	17.77	\pm	0.01	$&$	17.65	\pm	0.01	$&$	17.53	\pm	0.02	$&$	17.49	\pm	0.01	$\\
20.8744	&	115	&$	17.77	\pm	0.01	$&$	17.66	\pm	0.01	$&$	17.54	\pm	0.02	$&$	17.48	\pm	0.02	$\\
21.0717	&	115	&$	17.76	\pm	0.01	$&$	17.68	\pm	0.01	$&$	17.55	\pm	0.01	$&$	17.49	\pm	0.01	$\\
21.2673	&	115	&$	17.79	\pm	0.01	$&$	17.67	\pm	0.01	$&$	17.56	\pm	0.01	$&$	17.48	\pm	0.01	$\\
21.4580	&	115	&$	17.81	\pm	0.01	$&$	17.69	\pm	0.01	$&$	17.58	\pm	0.01	$&$	17.49	\pm	0.01	$\\
92.4295	&	701	&$	19.66	\pm	0.02	$&$	19.48	\pm	0.02	$&$	19.29	\pm	0.02	$&$	19.22	\pm	0.03	$\\
93.2890	&	679	&$	19.70	\pm	0.02	$&$	19.49	\pm	0.01	$&$	19.32	\pm	0.02	$&$	19.22	\pm	0.03	$\\
108.8565	&	686	&$	19.92	\pm	0.02	$&$	19.72	\pm	0.01	$&$	19.58	\pm	0.02	$&$	19.47	\pm	0.02	$\\
179.6620	&	1695	&$	20.78	\pm	0.02	$&$	20.53	\pm	0.02	$&$	20.31	\pm	0.02	$&$	20.30	\pm	0.03	$\\
189.9125	&	1714	&$	20.85	\pm	0.03	$&$	20.60	\pm	0.02	$&$	20.42	\pm	0.02	$&$	20.35	\pm	0.03	$\\
277.0450	&	1708	&$	21.55	\pm	0.05	$&$	21.22	\pm	0.04	$&$	21.04	\pm	0.04	$&$	21.01	\pm	0.06	$\\
363.9306	&	1697	&$	21.96	\pm	0.09	$&$	21.67	\pm	0.05	$&$	21.48	\pm	0.06	$&$	21.39	\pm	0.06	$\\
533.5294	&	1707	&$	22.43	\pm	0.08	$&$	22.07	\pm	0.06	$&$	21.91	\pm	0.06	$&$	21.94	\pm	0.08	$\\
959.0369	&	1709	&$	23.18	\pm	0.05	$&$	22.26	\pm	0.03	$&$	21.86	\pm	0.06	$&$	22.36	\pm	0.15	$\\
960.8429	&	1709	&$	23.21	\pm	0.05	$&$	22.27	\pm	0.03	$&$	21.92	\pm	0.05	$&$	22.29	\pm	0.12	$\\
3985.5129	&	3922	&$	23.81	\pm	0.08	$&$	22.86	\pm	0.05	$&$	22.36	\pm	0.06	$&$	22.61	\pm	0.09	$\\
4244.3423	&	1700	&$	23.87	\pm	0.11	$&$	23.10	\pm	0.06	$&$	22.65	\pm	0.08	$&$	22.68	\pm	0.17	$\\
4673.6840	&	1896	&$	23.90	\pm	0.19	$&$	23.16	\pm	0.12	$&$	22.76	\pm	0.13	$&$	22.78	\pm	0.19	$\\
29225.2102	&	4906	&$	24.12	\pm	0.08	$&$	23.28	\pm	0.05	$&$	22.88	\pm	0.07	$&$	23.38	\pm	0.18	$\\
\hline
\end{longtable}
\begin{list}{}{}
\item[$^{\mathrm{(a)}}$] Corrected for Galactic foreground reddening.
\end{list}

\begin{table*}
\caption{$JH$ photometric data}             
\label{JH}      
\centering                         
\begin{tabular}{r c c c}        
\hline\hline  
$T_\mathrm{mid} - T_0$ [ks] & Exposure [s] & \multicolumn{2}{c}{
Brightness$^{\mathrm{(a)}}$ mag$_\mathrm{AB}^{\mathrm{(b)}}$}  \\
 & & $J$ & $H$  \\
\hline 
3.6948	&	386	&$	16.37	\pm	0.02	$&$	16.30	\pm	0.03	$\\
4.2991	&	730	&$	16.45	\pm	0.02	$&$	16.33	\pm	0.03	$\\
5.0877	&	752	&$	16.51	\pm	0.02	$&$	16.43	\pm	0.03	$\\
5.7092	&	386	&$	16.52	\pm	0.02	$&$	16.45	\pm	0.03	$\\
6.1479	&	387	&$	16.56	\pm	0.02	$&$	16.48	\pm	0.03	$\\
6.5792	&	387	&$	16.60	\pm	0.02	$&$	16.51	\pm	0.03	$\\
7.0072	&	388	&$	16.63	\pm	0.02	$&$	16.53	\pm	0.03	$\\
7.4362	&	389	&$	16.65	\pm	0.02	$&$	16.53	\pm	0.03	$\\
7.8569	&	389	&$	16.69	\pm	0.02	$&$	16.59	\pm	0.03	$\\
8.2799	&	389	&$	16.73	\pm	0.02	$&$	16.60	\pm	0.03	$\\
8.7311	&	385	&$	16.75	\pm	0.02	$&$	16.64	\pm	0.03	$\\
9.1512	&	385	&$	16.77	\pm	0.02	$&$	16.69	\pm	0.03	$\\
9.5755	&	387	&$	16.83	\pm	0.02	$&$	16.72	\pm	0.03	$\\
9.9990	&	388	&$	16.81	\pm	0.02	$&$	16.76	\pm	0.03	$\\
10.4291	&	391	&$	16.84	\pm	0.02	$&$	16.75	\pm	0.03	$\\
10.8530	&	386	&$	16.90	\pm	0.02	$&$	16.77	\pm	0.03	$\\
11.2784	&	388	&$	16.91	\pm	0.02	$&$	16.77	\pm	0.03	$\\
11.7115	&	392	&$	16.90	\pm	0.02	$&$	16.77	\pm	0.03	$\\
12.1371	&	388	&$	16.90	\pm	0.02	$&$	16.80	\pm	0.03	$\\
12.5670	&	388	&$	16.94	\pm	0.02	$&$	16.86	\pm	0.03	$\\
12.9968	&	388	&$	17.00	\pm	0.02	$&$	16.88	\pm	0.03	$\\
13.4258	&	389	&$	16.98	\pm	0.02	$&$	16.88	\pm	0.03	$\\
13.8544	&	387	&$	17.01	\pm	0.02	$&$	16.88	\pm	0.03	$\\
14.2805	&	388	&$	17.04	\pm	0.02	$&$	16.88	\pm	0.03	$\\
14.7055	&	385	&$	17.07	\pm	0.02	$&$	16.93	\pm	0.03	$\\
15.1351	&	393	&$	17.05	\pm	0.02	$&$	16.93	\pm	0.03	$\\
15.5656	&	387	&$	17.08	\pm	0.02	$&$	16.98	\pm	0.03	$\\
15.9917	&	388	&$	17.10	\pm	0.02	$&$	17.00	\pm	0.03	$\\
16.4118	&	381	&$	17.12	\pm	0.02	$&$	17.05	\pm	0.03	$\\
17.2991	&	820	&$	17.16	\pm	0.02	$&$	17.03	\pm	0.03	$\\
18.0559	&	742	&$	17.22	\pm	0.02	$&$	17.11	\pm	0.03	$\\
18.8383	&	750	&$	17.26	\pm	0.02	$&$	17.15	\pm	0.03	$\\
19.6202	&	720	&$	17.28	\pm	0.02	$&$	17.21	\pm	0.03	$\\
20.3949	&	748	&$	17.29	\pm	0.02	$&$	17.26	\pm	0.03	$\\
21.1932	&	753	&$	17.34	\pm	0.02	$&$	17.24	\pm	0.03	$\\
92.4549	&	754	&$	19.04	\pm	0.09	$&$	18.83	\pm	0.10	$\\
93.3151	&	733	&$	19.09	\pm	0.09	$&$	18.90	\pm	0.09	$\\
107.4009	&	1751	&$	19.23	\pm	0.06	$&$	19.01	\pm	0.08	$\\
179.6887	&	1751	&$	20.09	\pm	0.09	$&$	19.88	\pm	0.11	$\\
189.9391	&	1770	&$	20.09	\pm	0.08	$&$	19.92	\pm	0.12	$\\
277.0709	&	1762	&$	20.54	\pm	0.32	$&$	20.54	\pm	0.23	$\\
363.9571	&	1750	&$	20.88	\pm	0.18	$&$	21.01	\pm	0.29	$\\
533.5557	&	1750	&$	21.53	\pm	0.32	$&$	>21.28			$\\
959.0635	&	1750	&$	>21.63			$&$	>21.18			$\\
3985.5371	&	3969	&$	21.74	\pm	0.28	$&$	>21.37			$\\
4244.2658	&	1750	&$	>21.69			$&$	>21.21			$\\
4673.7890	&	1750	&$	>21.37			$&$	>20.95			$\\
\hline
\end{tabular}
\begin{list}{}{}
\item[$^{\mathrm{(a)}}$] Corrected for Galactic foreground reddening. Converted
to AB magnitudes for consistency with Table \ref{griz}.
\item[$^{\mathrm{(a)}}$] For the SED fitting, the additional error of the absolute
calibration of 0.05 mag was added.
\end{list}
\end{table*}

\begin{table*}
\caption{Secondary standards in the GRB field in the GROND
filter bands used for the calibration}             
\label{standards091127}      
\centering                         
\begin{tabular}{c c c c c c c c}        
\hline\hline  
Star & R.A., Dec & $g'$ & $r'$ & $i'$ & $z'$ & $J$ & $H$ \\
 number & [J2000] & (mag$_\mathrm{AB}$) & (mag$_\mathrm{AB}$) & (mag$_\mathrm{AB}$) & (mag$_\mathrm{AB}$) &
 (mag$_\mathrm{Vega}$) & (mag$_\mathrm{Vega}$) \\
\hline                     
1 & 02:26:21.05, $-$18:57:19.1 & $15.18\pm0.03$ & $14.49\pm0.03$ & $14.22\pm0.03$ & $14.07\pm0.03$ & $13.03\pm0.05$ & $12.57\pm0.05$\\
2 & 02:26:12.17, $-$18:57:17.6 & $17.48\pm0.03$ & $16.64\pm0.03$ & $16.26\pm0.03$ & $16.06\pm0.03$ & $14.47\pm0.05$ & $14.38\pm0.05$\\
3 & 02:26:12.14, $-$18:57:02.9 & $17.74\pm0.03$ & $16.96\pm0.03$ & $16.80\pm0.03$ & $16.71\pm0.03$ & $14.93\pm0.05$ & $15.35\pm0.05$\\
4 & 02:26:23.64, $-$18:58:17.8 & $22.17\pm0.03$ & $20.43\pm0.03$ & $19.34\pm0.03$ & $18.80\pm0.03$ & - & - \\
5 & 02:26:25.03, $-$18:58:45.5 & $20.59\pm0.03$ & $19.05\pm0.03$ & $18.16\pm0.03$ & $17.71\pm0.03$ & - & - \\
\hline                                   
\end{tabular}
\end{table*}

\end{document}